%% file: composite_lag_rv2.tex
\documentclass[twocolumn,iop]{emulateapj}
\usepackage{apjfonts}
\usepackage{color}

\usepackage{amsmath,graphicx,longtable,hyperref}
\usepackage{natbib,threeparttable,rotating}
\usepackage{ulem}

\newcommand{\etal}{et al.}
\newcommand{\hbeta}{H{$\beta$}}
\newcommand{\halpha}{H{$\alpha$}}

\newcommand{\CIV}{C\,{\sevenrm IV}}

\newcommand{\CIII}{C\,{\sevenrm III]}}

\def\FeII{Fe\,{\sc ii}}
\def\MgII{Mg\,{\sc ii}}
\def\HeII{He\,{\sc ii}}

\newcommand{\bracket}[1]{\left\langle#1\right\rangle}
   \font\sevenrm=cmr7 scaled 1000
\newcommand{\comments}[1]{}

\def\kms{{\rm km\,s^{-1}}}

\begin{document}

\title{The Sloan Digital Sky Survey Reverberation Mapping Project: Composite Lags at $z\lesssim 1$}

\author{Jennifer I-Hsiu Li$^{1}$, Yue Shen$^{1,2,*}$, Keith Horne$^3$, W.~N. Brandt$^{4,5,6}$, Jenny E. Greene$^7$, C.~J. Grier$^{4,5}$, Luis C. Ho$^{8,9}$, Chris Kochanek$^{10}$, Donald P. Schneider$^{4,5}$, Jonathan R.~Trump$^{11}$, Kyle S. Dawson\altaffilmark{12}, Kaike Pan\altaffilmark{13}, Dmitry Bizyaev\altaffilmark{13,14}, Daniel Oravetz\altaffilmark{13}, Audrey Simmons\altaffilmark{13}, Elena Malanushenko\altaffilmark{13}}


\altaffiltext{1}{Department of Astronomy, University of Illinois at Urbana-Champaign, Urbana, IL 61801, USA}
\altaffiltext{2}{National Center for Supercomputing Applications, University of Illinois at Urbana-Champaign, Urbana, IL 61801, USA}
\altaffiltext{*}{Alfred P. Sloan Research Fellow}
\altaffiltext{3}{SUPA Physics/Astronomy, Univ. of St. Andrews, St. Andrews KY16 9SS, Scotland, UK}
\altaffiltext{4}{Department of Astronomy \& Astrophysics, The Pennsylvania State University, University Park, PA, 16802, USA}
\altaffiltext{5}{Institute for Gravitation and the Cosmos, The Pennsylvania State University, University Park, PA 16802, USA}
\altaffiltext{6}{Department of Physics, 104 Davey Lab, The Pennsylvania State University, University Park, PA 16802, USA}
\altaffiltext{7}{Department of Astrophysical Sciences, Princeton University, Princeton, NJ 08544, USA}
\altaffiltext{8}{Kavli Institute for Astronomy and Astrophysics, Peking University, Beijing 100871, China}
\altaffiltext{9}{Department of Astronomy, School of Physics, Peking University, Beijing 100871, China}
\altaffiltext{10}{Department of Astronomy, The Ohio State University, 140 West 18th Avenue, Columbus, OH 43210, USA}
\altaffiltext{11}{Department of Physics, University of Connecticut, 2152 Hillside Rd Unit 3046, Storrs, CT 06269, USA}
\altaffiltext{12}{Department of Physics and Astronomy, University of Utah, Salt Lake City, UT 84112, USA}
\altaffiltext{13}{Apache Point Observatory and New Mexico State University, P.O. Box 59, Sunspot, NM, 88349-0059, USA}
\altaffiltext{14}{Sternberg Astronomical Institute, Moscow State University, Moscow}

\shorttitle{SDSS-RM: Composite Lags at $z\lesssim 1$}
\shortauthors{Li \etal}

\begin{abstract}

We present composite broad-line region (BLR) reverberation-mapping lag measurements for \halpha, \hbeta, \HeII\,$\lambda4686$ and \MgII\ for a sample of 144,  $z\lesssim 1$ quasars from the Sloan Digital Sky Survey Reverberation Mapping (SDSS-RM) project. Using only the 32-epoch spectroscopic light curves in the first 6-month season of SDSS-RM observations, we compile correlation-function measurements for individual objects and then coadd them to allow the measurement of the average lags for our sample at mean redshifts of $0.4$ (for \halpha) and $\sim 0.65$ (for the other lines). At similar quasar luminosities and redshifts, the sample-averaged lag decreases in the order of \MgII, \halpha, \hbeta\ and \HeII. This decrease in lags is accompanied by an increase in the mean line width of the four lines, and is roughly consistent with the virialized motion for BLR gas in photoionization equilibrium. These are among the first RM measurements of stratified BLR structure at $z>0.3$. Dividing our sample by luminosity, \halpha\ shows clear evidence of increasing lags with luminosity, consistent with the expectation from the measured BLR size-luminosity relation based on \hbeta. The other three lines do not show a clear luminosity trend in their average lags due to the limited dynamic range of luminosity probed and the poor average correlation signals in the divided samples, a situation that will be improved with the incorporation of additional photometric and spectroscopic data from SDSS-RM. We discuss the utility and caveats of composite-lag measurements for large statistical quasar samples with reverberation-mapping data.
\end{abstract}

\section{Introduction}\label{sec:intro}

Reverberation mapping (RM) is a technique used to infer the size of the broad-line region (BLR) in active galactic nuclei (AGN) and quasars by measuring the time delay between the continuum and broad-line flux variations \citep[e.g.,][]{Blandford_McKee_1982, Peterson_1993}. Combining the BLR size with the virial velocity inferred from the width of the broad lines, one can derive a virial estimate of the mass of the black hole. This is the primary technique used to measure BH masses in AGN and quasars, and anchors secondary methods of active BH mass estimation based on single-epoch spectroscopy \citep[for a recent review, see,][]{Shen_2013}. 

Over the past two decades, RM measurements have been performed for dozens of low-redshift ($z<0.3$) AGN and quasars \citep[e.g.,][]{Peterson_etal_1998a,Kaspi_etal_2000,Kaspi_etal_2005,Peterson_etal_2002,
Peterson_etal_2004,Bentz_etal_2009,Bentz_etal_2010,Bentz_etal_2013,Denney_etal_2009a,Denney_etal_2010,Rafter_etal_2011,Rafter_etal_2013, Barth_etal_2011a,Barth_etal_2011b,Barth_etal_2013,Barth_etal_2015,Grier_etal_2012,Du_etal_2014,Du_etal_2015,Du_etal_2016,Hu_etal_2015}, and the feasibility and potential of this technique for measuring BH masses and understanding the inner structure of AGN and quasars has been well demonstrated. In recent years, RM has been attempted at higher redshifts ($z>0.3$) and for high-luminosity quasars as well \citep[e.g.,][]{Kaspi_etal_2007,Trevese_etal_2014, Jiang_etal_2016}, but successful measurements are still rare in these regimes. An interesting new approach to RM is to use multi-object spectrographs to study hundreds of quasars simultaneously \citep[MOS-RM, e.g.,][]{Shen_etal_2015a,King_etal_2015}. In addition to the much improved observational efficiency, MOS-RM programs aim to detect time lags for uniformly selected samples of AGN and quasars at substantially higher redshifts and luminosities with multi-year time baselines to sample the slow variability patterns of these high-redshift objects. 

We are conducting one of the first MOS-RM programs using the SDSS Baryon Oscillation Spectroscopic Survey (BOSS) spectrograph \citep{Dawson_etal_2013,Smee_etal_2013} on the 2.5-m SDSS telescope \citep{Gunn_etal_2006}, accompanied by dedicated photometric monitoring using a number of ground-based wide-field imagers. Details of the SDSS-RM project are presented in \citet{Shen_etal_2015a}. Analyses of the initial season (2014) of spectroscopic data have led to the first robust detections of BLR lags at $z>0.3$, and a subset of 15 individual detections were reported by \citet[][]{Shen_etal_2016a}. 

The large sample of objects with RM data from the SDSS-RM project also enables an investigation of composite lag detections, where one may boost the signal-to-noise ratio (SNR) of the lag detection (inferred from the cross-correlation between continuum and line light curves) by stacking the results from individual objects \citep[e.g.,][]{Fine_etal_2012, Fine_etal_2013}. As illustrated by these authors, this technique is useful for measuring the {\it average} lag for a sample of quasars, even if the individual light curves are of insufficient quality to measure a reliable lag or if the correlated variability is buried in the random, uncorrelated intrinsic variability of each light curve. Composite lag measurements provide a complementary approach to individual lag measurements to perform RM studies for high-$z$ samples. 

In this work we test the feasibility of composite lag detections using a subset of SDSS-RM data from our first-season observations conducted in 2014. We use the well-calibrated first-season (6-month) spectroscopic light curves \textit{alone} to create the coadded correlation function. This slightly differs from \citet{Fine_etal_2013} which combined high-cadence continuum light curves with a few spectroscopic epochs, but the spirit is the same in both studies. This approach is carried out in parallel to our ongoing effort on individual lag detections, and it is particularly useful for lag detections with weak broad lines (such as \HeII). {The motivations for performing such an exercise with the SDSS-RM data are the following: (1) with composite lags we will attempt to measure average lags for different line species, in particular the weak broad line \HeII\,$\lambda4686$, in the same sample of quasars, allowing us to explore the stratified structure of quasar BLRs; (2) ongoing and future MOS-RM programs will produce light-curve data for large statistical quasar samples. Composite lag measurements then offer a promising way to boost the signal-to-noise ratio, and to measure the average lags for quasars binned by different physical properties. This work serves as a demonstration of concept for this approach using the first-season SDSS-RM spectroscopic-only data. Nevertheless, the lag measurements we present here are at mean redshifts $>0.3$, where RM measurements are rare, and include the first measurement of (composite) \HeII\ lags at such high redshifts. }

This paper is organized as follows. In \S\ref{sec:data} we describe the sample and light-curve data used, and in \S\ref{sec:tech} we describe the technical details of stacked lag detection, with several tests to demonstrate its robustness in \S\ref{sec:tests}. We present our results in \S\ref{sec:results}, and summarize our findings in \S\ref{sec:sum}, with an outlook for future improvement. Throughout the paper we adopt a flat $\Lambda$CDM cosmology with $\Omega_\Lambda=0.7$ and $H_{0}=70\,\kms\,{\rm Mpc}^{-1}$ in calculating luminosities. Unless stated otherwise, all stacked lag measurements are performed in the observed frame for the reasons discussed in \S\ref{sec:tech}. 

\section{Data}\label{sec:data}

SDSS-RM simultaneously monitors 849 broad-line quasars at $0.1<z<4.5$ with a flux limit of $i_{\rm psf}=21.7$ \citep{Shen_etal_2015a}. The spectroscopic data used in this work are from the 32 epochs taken in 2014 (from January to July) as part of the SDSS-RM project within the SDSS-III \citep{Eisenstein_etal_2011} Baryon Oscillation Spectroscopic Survey \citep[BOSS,][]{Dawson_etal_2013}. The wavelength coverage of BOSS spectra is $\sim 3650-10,400~\textrm{\AA}$, with a spectral resolution of $R\sim 2000$. Each epoch had a typical exposure time of 2 hrs, resulting a typical SNR of $\sim 4.5$ per $69\,\kms$ pixel at $g_{\rm psf}=21.2$ averaged over the $g$ band. The epoch-by-epoch spectra were pipeline-processed as part of the SDSS-III Data Release 12 \citep{dr12}, followed by a custom flux calibration scheme and improved sky subtraction as described by \citet{Shen_etal_2015a}. The improved spectrophotometry has a nominal absolute accuracy of $\sim 5\%$. 

We then performed a spectral-refinement procedure, designated ``PrepSpec'', on the custom flux-calibrated multi-epoch spectra, as described in detail in \citet{Shen_etal_2016a}. PrepSpec further improves the relative flux measurements by adjusting the flux levels of the individual epochs using the fluxes of narrow emission lines, assumed to remain constant over the monitoring period. The PrepSpec procedure derived continuum and broad-line light curves for each object in our sample, as well as measurements of the emission line widths from both the mean and the root-mean-squared (rms) spectra. These {\it spectroscopic-only} light curves form the basis of our composite-lag analysis. 

\begin{figure}
\centering
    \includegraphics[width=8.7cm]{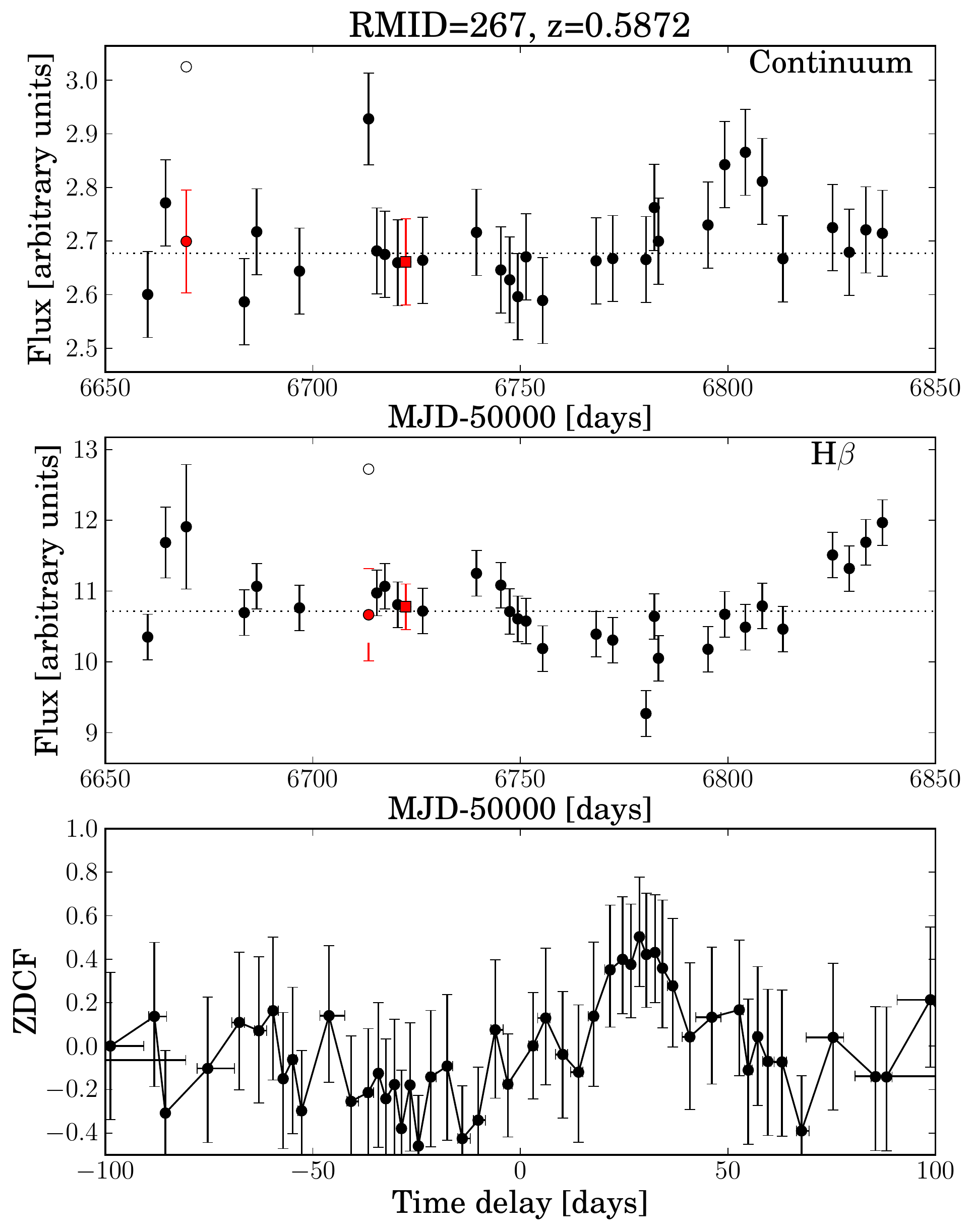}
    \caption{An example of the light curve adjustment process described in \S\ref{sec:data}. The top two panels show the continuum (estimated at rest-frame 5100\,\AA) and broad-line light curves. The open points are the epochs we identified as outliers, and the red points are the ``fixed'' light curve data (circles are for outliers and squares are for missing data). The bottom panel shows the ZDCF. This object is one of the 15 first-lag objects reported in \citet{Shen_etal_2016a} where a lag detection is possible with the spectro-only LCs.}
    \label{fig:lc}
\end{figure}

Occasionally a particular epoch will appear as an outlier in the light curve. These events are usually caused by the poor SNR of that particular epoch \citep[e.g., Epoch 3 and Epoch 7; see][]{Shen_etal_2016a}, but occasionally may be due to unknown systematics in the PrepSpec procedure. To build a uniform set of light-curve data, which is crucial when carrying on to further correlation function and coadding calculations, we identify outlier epochs with differences larger than 3 standard deviations from the linear interpolations of the original light curves and replace their flux measurements with linear interpolations from the rest of the light curve. Similarly, we interpolate the light curve at rare epochs with missing data (mostly due to bad spectrograph fibers) and assign the flux error to be the average of the remaining data points. The fraction of these ``fixed'' light curve data points is {9.7\%} of total light curve points. Figure \ \ref{fig:lc} presents an example of our light curve adjustment. With this adjustment, all objects have exactly the same cadence in their light curves, allowing a uniform binning of the correlation functions. 

As discussed in \citet{Shen_etal_2016a}, one concern of our analysis is that PrepSpec may underestimate the light-curve uncertainties, and an empirical upper limit on the flux uncertainties is 3\%. We have tested our calculations with the original PrepSpec errors and the inflated light curve errors, and found consistent results. To be conservative with the light-curve errors, all analyses presented in this paper will adopt the 3\% inflated errors. 

\input{table_lag_efix}

\section{Composite Lag Measurements}\label{sec:tech}

In this work we focus on the low-$z$ subset of our sample for which a lag should be detectable given the 6-month observed period of our first-season monitoring. Specifically we require that a continuum light curve at rest-frame 5100\,\AA\ is available, which is the commonly adopted reference continuum for low-$z$ RM work. This requirement limits our sample to $z\lesssim 1$ and 190 objects. We consider four broad lines that are of primary interest in this redshift regime and for which we have available light curves from PrepSpec: \halpha, \hbeta, \HeII\,$\lambda4686$, and \MgII. 

Furthermore, we remove objects whose expected observed-frame \hbeta\ lags \citep[following the $R-L$ relation in ][]{Bentz_etal_2013} exceed the maximum lag range searched for correlation ($\sim 120$\,days), given the total baseline of $\sim 180$ days of our light curves. These objects will not contribute correlation signals to the coadded correlation function, but will add noise instead. In addition, these objects are typically at higher luminosity and redshift, and thus will bias the sample-averaged luminosity upwards. Indeed, we have tested including these objects in the coadded correlation functions, and found the resulting average lags are almost identical, but the sample-averaged luminosity is higher compared to the results excluding them. This additional requirement reduces the total numbers of objects used in this study to 144 (roughly $\sim$25\% of sources are removed).

Table \ref{tab:lag} lists the composite-lag measurements for these four lines and the corresponding number of objects in each coadd. 

\subsection{Methodology of composite lags}\label{sec:method}

To measure the cross-correlation signal in individual objects, we use the $z$-transformed discrete correlation function \citep[ZDCF,][]{Alexander_2013}. ZDCF is a more robust method to measure time lags for sparse and unevenly-sampled light curves than the traditional discrete correlation function \citep[DCF,][]{Edelson_Krolik_1988} and the interpolated cross-correlation function \citep[ICCF,][]{Gaskell_Peterson_1987}. By using equal population binning and Fisher's $z$-transform, ZDCF provides accurate estimates for light curves with as few as $\sim 15$ epochs. Compared with the ICCF and DCF, ZDCF provides more reliable error estimation (whereas ICCF does not provide errors on the correlation), which is useful in weighting the data points in the coadded correlation; more importantly, {unlike the ICCF, the ZDCF does not suffer from correlated errors in the continuum and broad-line light curves as measured from the same spectrum}. Finally, given the uniform sampling of light curves used here, the automatic time-lag binning of ZDCF yields the same binning for all objects, allowing a straightforward coaddition of points in each time-lag bin. These properties of the ZDCF make it the ideal tool to measure the average time lag for a given sample in our database. The bottom panel of Figure\ \ref{fig:lc} shows the ZDCF for an example object in our sample.

One drawback of using the ZDCFs in the stacking is that the individual cross correlations are stacked in the observed frame (unless some sort of interpolation and rebinning of the individual ZDCFs is used). Therefore there is an additional broadening of the stacked ZDCF due to different time-dilation factors in different objects. Fortunately, as further discussed in \S\ref{sec:result1}, when scaled by the sample-averaged $(1+z)$ factor, observed-frame coadded lags (without rebinning) are consistent with the rebinned rest-frame coadded lags, without the complications of data interpolation. Given the many advantages of ZDCFs discussed above, we will use ZDCFs and the observed-frame stacks to demonstrate our methodology below. 

Once we have measured the individual ZDCF$_{i,j}$ for object $i$ and time-lag bin $j$, and the associated errors $\sigma_{i,j}$, we create a composite ZDCF in two different ways. The first approach is a simple median coadd, where we take the median of the ZDCF distribution in each time-lag bin. In the second approach, we assign the individual ZDCF points in each time-lag bin an object-based weight
\begin{equation}
W_{i}={\rm median}\left[\frac{1}{\sigma_{i,*}^2}\right]\ ,
\end{equation}
and calculate the weighted mean and uncertainty in each time-lag bin. In this definition, the weight is on an object-by-object basis and is identical across all time-lag bins for the same object.\footnote{We adopt this weighting scheme instead of using different weights $1/\sigma^2_{i,j}$ for the same object across time-lag bins to avoid ``glitches'' in the $1/\sigma^2_{i,j}$ distribution due to unknown systematics in our light curves. The individual-epoch weighting scheme can lead to large fluctuations in the coadded ZDCF across time-lag bins. } The weighted mean provides a better measurement of the coadded ZDCF by up-weighting signals from high-quality light curves, and we take these results based on the weighted mean as our fiducial results. Conversely, both the simple-median approach and the weighted-mean approach account for contributions from all objects and hence the resulting coadded correlation is not dominated by a few objects. 

To quantify the inferred {\it average} lags, we use a second-order polynomial to iteratively fit the coadded ZDCF points within $\pm 50$\ days around the model peak until the peak position converges to within 10\% of itself. We also measure a centroid by directly using the coadded ZDCF data points {within $\pm 50$\ days of the best-fit polynomial peak} and found consistent results. Bootstrap resampling of the objects contributing to the coadded ZDCFs is adopted to estimate the values and uncertainties in the measured average lags. We adopt the median of the lag distribution from the bootstrap samples as the measured lag to provide more robust estimates than using the peak/centroid measured from the original coadded ZDCF. 

To assess visually the statistical significance of the coadded lag detection, we generate mock light curve pairs by shuffling the real light curve epochs for each object and measure the coadded ZDCF from the individual ZDCFs derived from the mock data. This Monte Carlo procedure was repeated for 100 realizations, and the median and the 16th and 84th percentiles of the distribution of the coadded ZDCF are recorded as the expected 1$\sigma$ range of coadded signals expected from uncorrelated light curves. By randomly shuffling the real light curve epochs we destroy any intrinsic correlation.\footnote{This shuffled-epoch test \citep[e.g.,][]{Shen_etal_2016a} also removes the characteristic red noise of quasar variability, but such variability will not introduce correlated continuum and line variability on the timescales of interest here. } For well detected composite lags, the coadded ZDCF should lie above the expected uncertainties from uncorrelated light curves.



As a consistency check, we coadded the ICCFs of individual objects and found consistent signals. However, there is often extra signal near zero lag due to correlated errors between the continuum and broad-line light curves \citep[see discussions in ][]{Shen_etal_2016a}. The calculation of the ZDCF avoids such complications, and provides a cleaner coadded correlation. 

A coadded correlation function is a diluted and broadened version of the individual correlation functions because different objects have different intrinsic lags and redshift dilation, and because low-variability light curves contribute both signal and noise. If appropriately weighted, this averaging process should reduce the noise from individual correlation measurements, and boost the SNR of the average lag enough to make a detection possible. We intentionally use all available objects so that the average lag represents the entire quasar sample, even though the inclusion of the low-variability subset of the sample may lead to noisier measurements.

In addition to coadding all objects, we also coadded the ZDCF for all sources excluding the 15 first-lag sources reported in \citet{Shen_etal_2016a} (the ``other'' sample). For \halpha\ and \hbeta, we are able to further divide the sources into low and high luminosity bins since the coadded ZDCF peaks are more significant for these two lines. We calculated 16 sets of coadded ZDCF in total, and the results are presented in Table \ref{tab:lag} and discussed in \S\ \ref{sec:results}.

Although our default results are based on the coadded ZDCF in the observed frame, we also compute average lags from the coadded ZDCF in the rest-frame of the quasars with rebinning and interpolation of the light curves. The results based on the rest-frame coadds are listed in the last two columns of Table \ref{tab:lag}.

\begin{figure}
\centering
	\includegraphics[width=8.7cm]{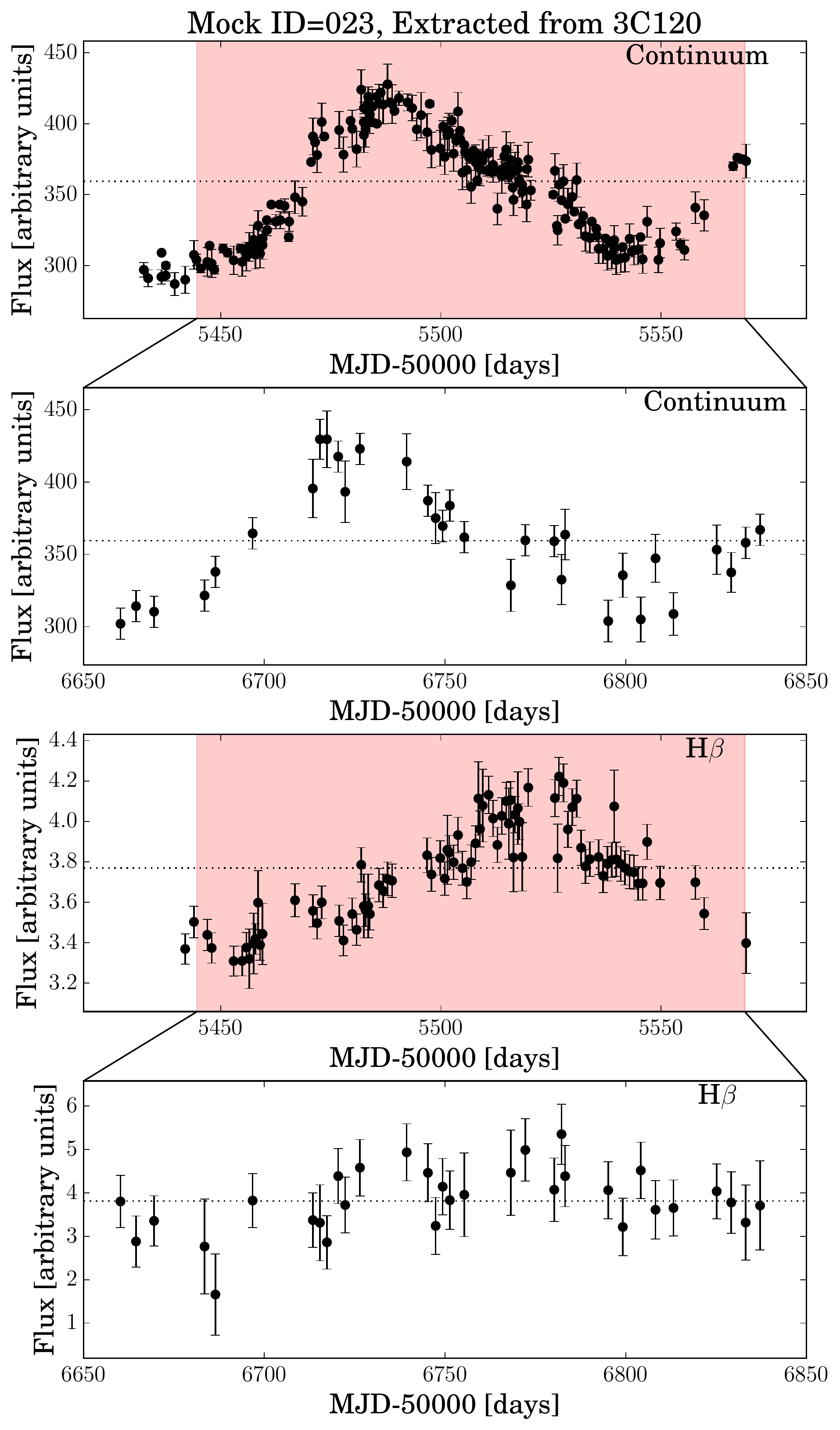}
    \caption{An example of a mock LC generated as described in \S\ref{sec:localRM}. From the top to bottom panels are the original continuum LC, mock continuum LC, original line LC and mock line LC. The red shaded areas show the segments of the original LCs that are used to generate the mock LCs.}
    \label{fig:mock}
\end{figure}

\subsection{The robustness of composite lags}\label{sec:tests}

Before we present the composite lag results for our SDSS-RM sample, we perform a series of additional tests to demonstrate the feasibility and robustness of this coadding technique.

\subsubsection{Test on the local RM sample}\label{sec:localRM}
The first test is to perform exactly the same coadding technique as described in \S\ref{sec:method} on the \citet{Bentz_etal_2013} sample, which includes 71 pairs of LCs from 41 local sources. This is a crucial test to evaluate the reliability of the coadding technique with a sample of known lags.

To match our sample size and observing period, we first generate 144 pairs of mock LCs by randomly choosing LC segments of individual objects from the \citet{Bentz_etal_2013} sample, which are assigned redshifts drawn from the redshift distribution of SDSS-RM quasars. The LC segments are required to span 180 days (in the observed-frame with the assigned redshift) with both continuum and line observations. We then degrade the mock LCs by increasing the measurement errors such that the RMS variability of the LCs (normalized by measurement errors) matches that of the SDSS-RM sample on average. This is because the local RM AGN typically have larger variability amplitudes than SDSS-RM quasars at higher luminosities. Finally, we match the exact cadence of SDSS-RM on the mock LCs with linear interpolation. See Fig.\ \ref{fig:mock} for an example of the original and degraded LCs. 

For sources with short observing durations and high cadences (i.e., corresponding to short lags), we deliberately assign high redshifts ($z>1$) to ensure they can be observed with our observing cadence. For the few sources with long lags or sparse LCs, we do not include them in our test as the lags cannot be detected within the period of 180 days. 51 out of the 71 LC pairs are used for generating the mock LCs and the coadding. The resulting sample of mock LCs has a similar size, cadence and variability/noise ratio to our SDSS-RM sample. The mock LCs are of too poor quality to formally detect a lag in individual objects.

\begin{figure}
\centering
	\begin{tabular}{@{}cc@{}}
	\includegraphics[width=0.48\textwidth]{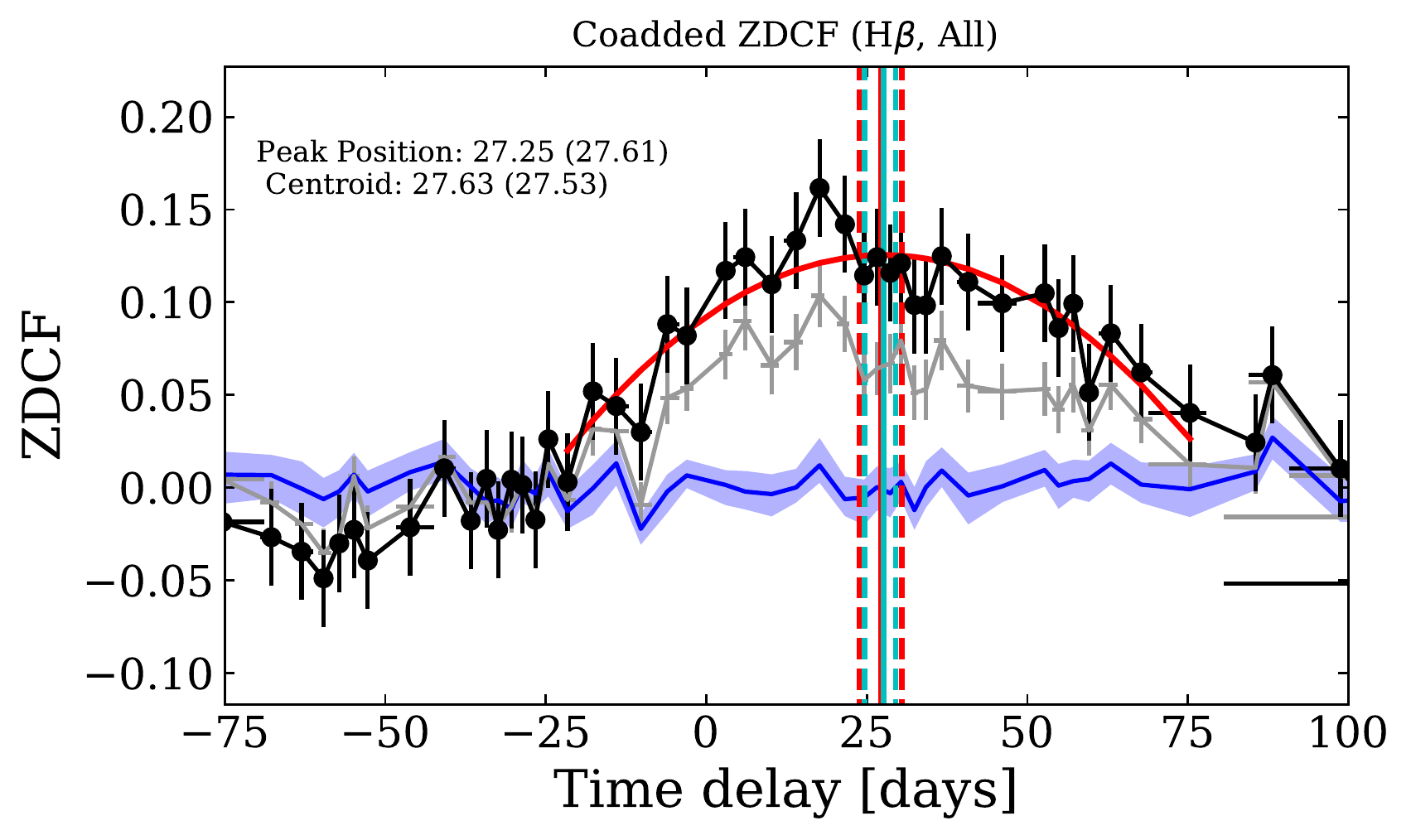} \\
        \includegraphics[width=0.48\textwidth]{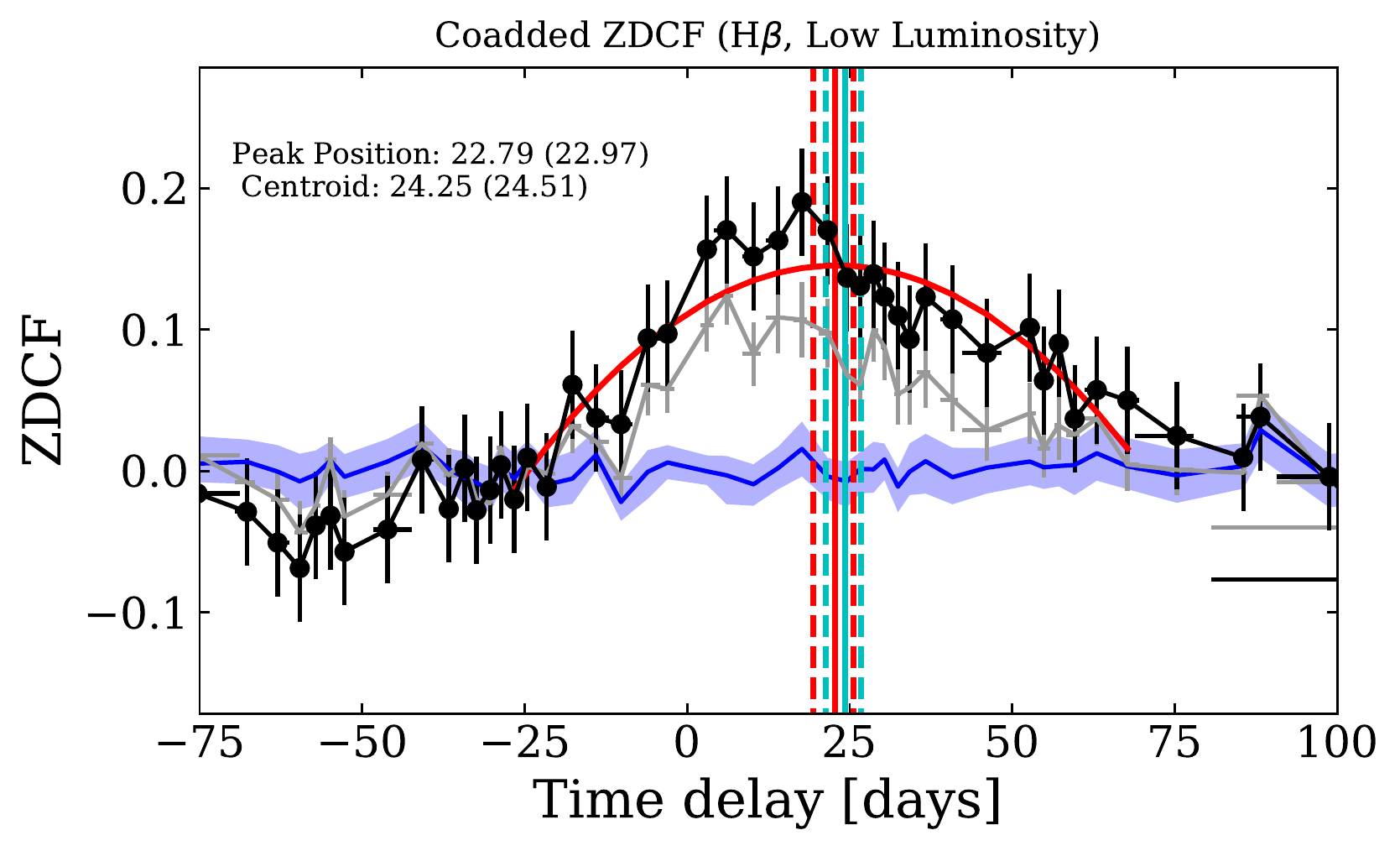} \\
        \includegraphics[width=0.48\textwidth]{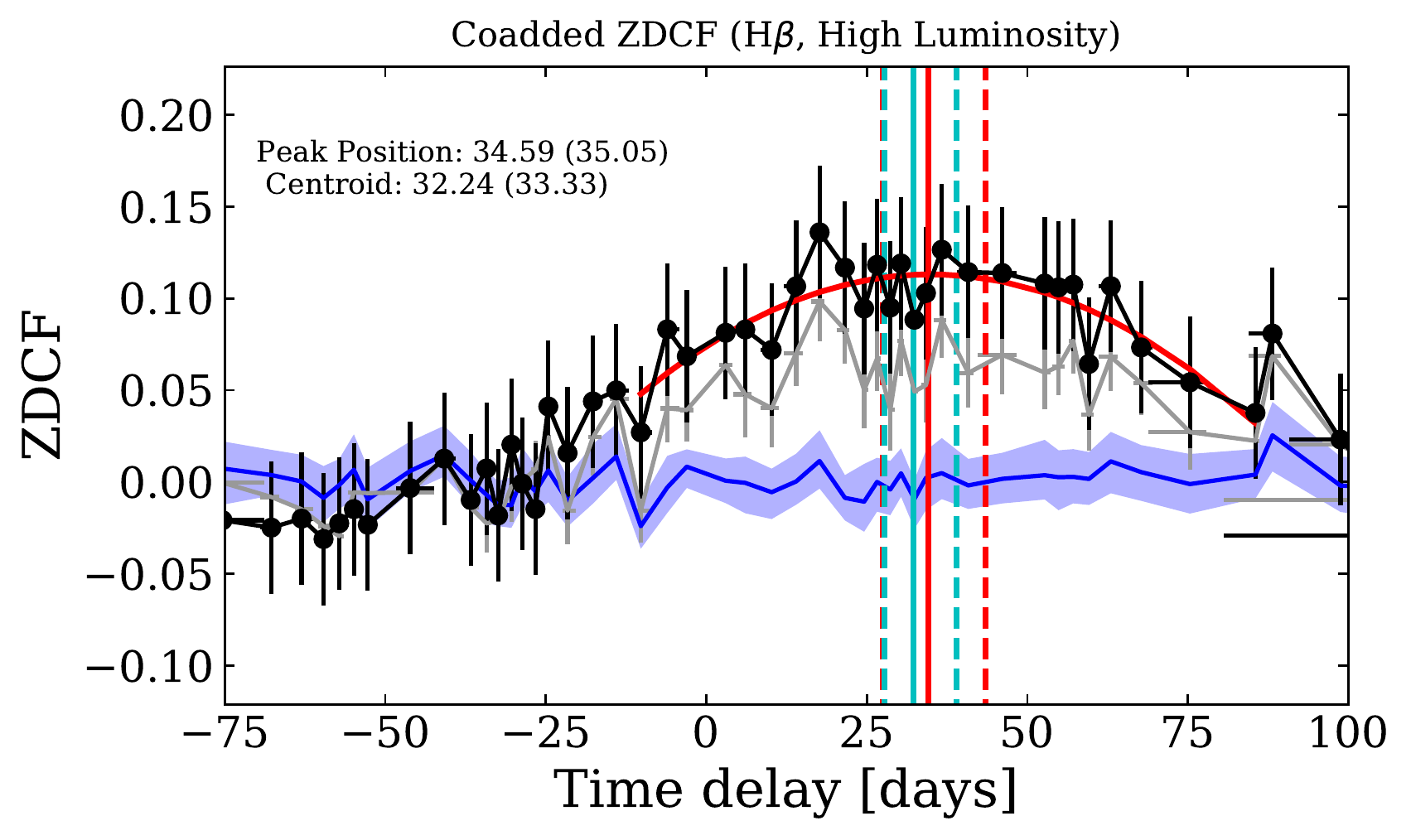}
	\end{tabular}
    \caption{Coadded ZDCF using the LCs from the \citet{Bentz_etal_2013} sample, following the methodology detailed in \S\ref{sec:method}. From top to bottom, the panels show the full sample, and the low and high luminosity subsets. In each panel, the blue line and the shaded band are the expected signal and uncertainty (16th and 84th percentiles of the distribution) generated from mock light curves with no intrinsic lags. The black data points show the coadded ZDCF with the weighted mean. Red curves are the polynomial fits to the coadded ZDCF within $\pm$ 50 days around the peak. Bootstrap resampling is used to estimate the uncertainty in the peak measurements. The estimated peak of the correlation (median of the bootstrap distribution) and its $1\sigma$ uncertainties are indicated by the red (polynomial fit) and cyan (direct centroid) vertical lines. The values measured from the median of the bootstrap distribution are marked in the upper-left corner, with the values measured from the original ZDCF indicated in the parentheses. For comparison, the grey line shows the coadded ZDCF with the simple median approach }
    \label{fig:bentz_coadd}
\end{figure}

\begin{figure}
\centering
    \includegraphics[width=0.48\textwidth]{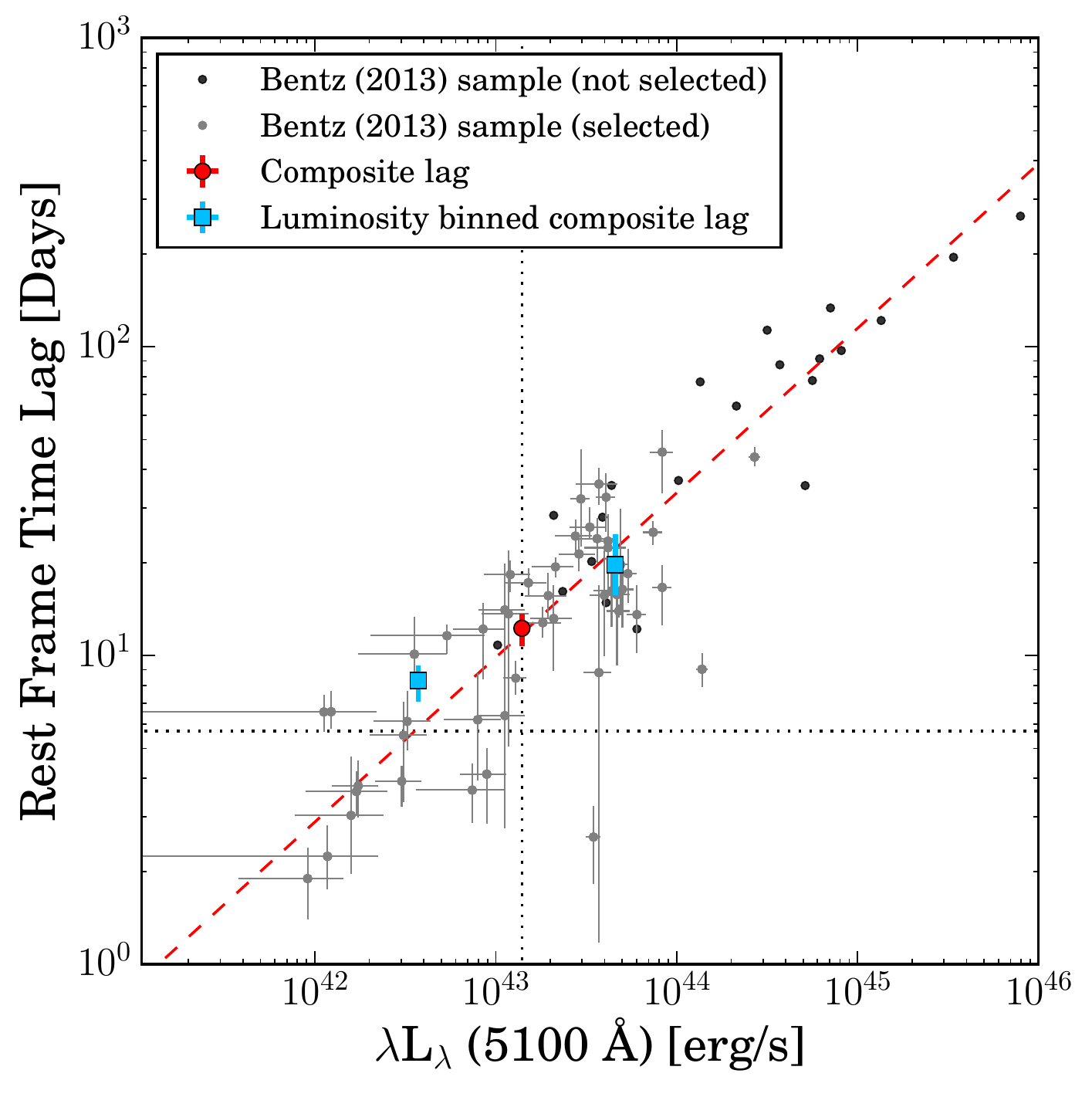} 
    \caption{\hbeta\ BLR radius and luminosity scatter plot from \citet{Bentz_etal_2013}. The grey dots shows the data points used for generating mock LCs in our test and the black dots were excluded due to insufficient coverage or lags exceeding our baseline. The red dashed line is the $R-L$ relation derived by \citet{Bentz_etal_2013}. The red circle and the cyan squares are the average lags of the whole and luminosity-divided samples calculated by the coadding technique, with negligible error bars. The average cadence of the SDSS-RM observation and the median luminosity of the mock sample are indicated by the horizontal and vertical dotted lines, respectively.}
    \label{fig:lag_bentz}
\end{figure}

We then perform the coadding technique as described previously in \S\ref{sec:method} on all mock LCs and two luminosity-divided subsets. The results are shown in Fig. \ref{fig:bentz_coadd}. Objects in the \citet{Bentz_etal_2013} sample have larger variability than SDSS-RM objects, and the coadded ZDCF yields a well-defined peak. We show the coadded results (after de-redshifting to the rest-frame) on the original $R-L$ plot from \citet{Bentz_etal_2013} in Fig. \ref{fig:lag_bentz}. The results agree with the \citet{Bentz_etal_2013} $R-L$ relation nicely for the full sample and the luminosity-divided samples. It is worth noting that the coadded lags are measured from the coadded correlation function, rather than from the simple average of individual lags, which are usually difficult to measure in low-quality LCs.

This test demonstrates that the coadding technique can yield consistent, sample-averaged lags. 

\subsubsection{SNR Dependence}

We further test the robustness of the coadding technique at different SNR of the light curves. Again, we start from the mock sample generated from the \citet{Bentz_etal_2013} sample as described in \S\ref{sec:localRM}. Next, we further degrade the sample SNR by applying a constant inflating factor to the flux errors of all the LCs in the sample (but keeping the original LC flux unchanged). These inflated LC errors are propagated to the calculation of the individual ZDCF. This way, the individual ZDCF will have lower significance and larger uncertainties. Finally, we measure the coadded ZDCF and lags on the degraded samples following the same methodology as before. 

\begin{figure}
\centering
	\begin{tabular}{@{}cc@{}}
	\includegraphics[width=0.48\textwidth]{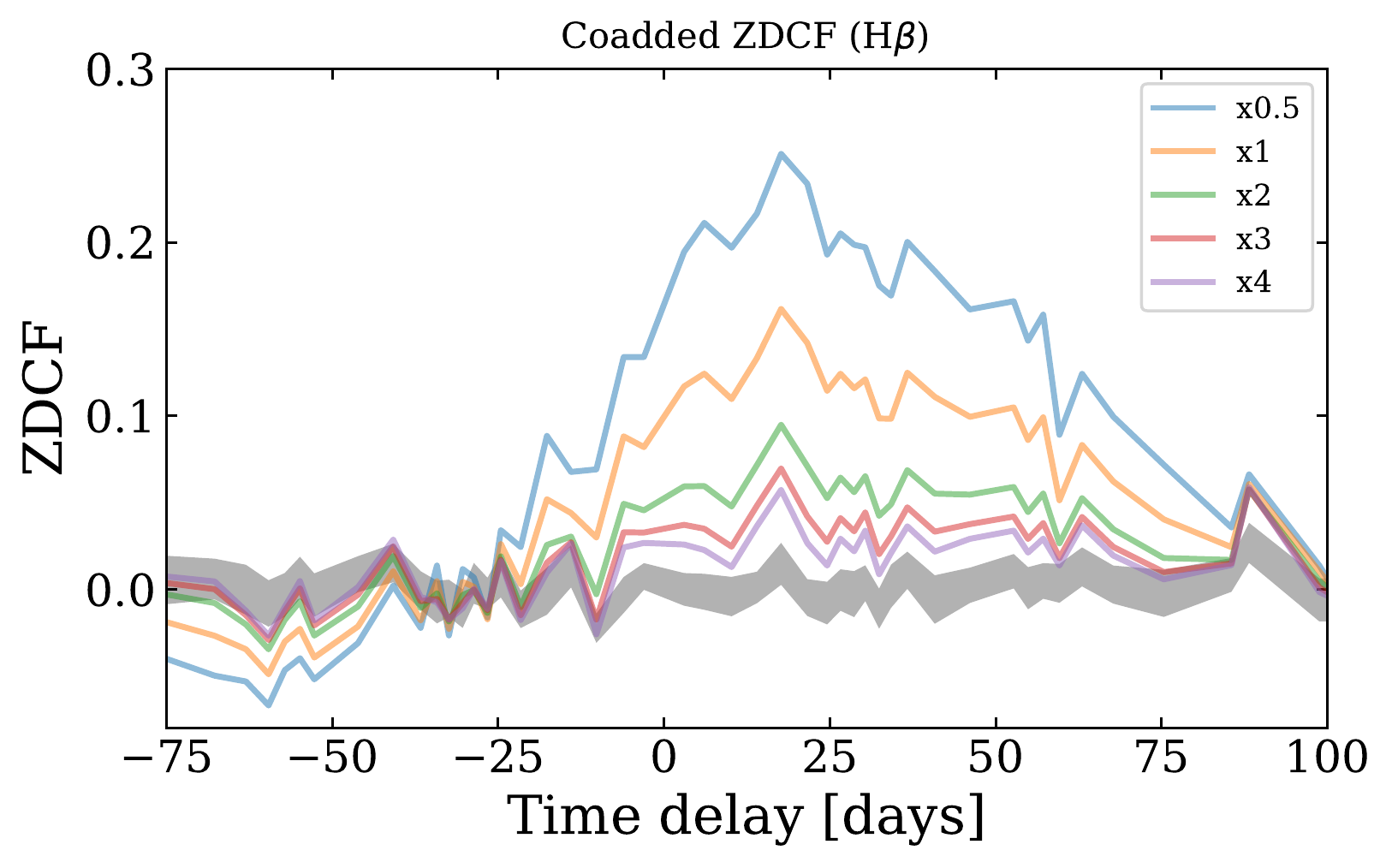} \\
        \includegraphics[width=0.48\textwidth]{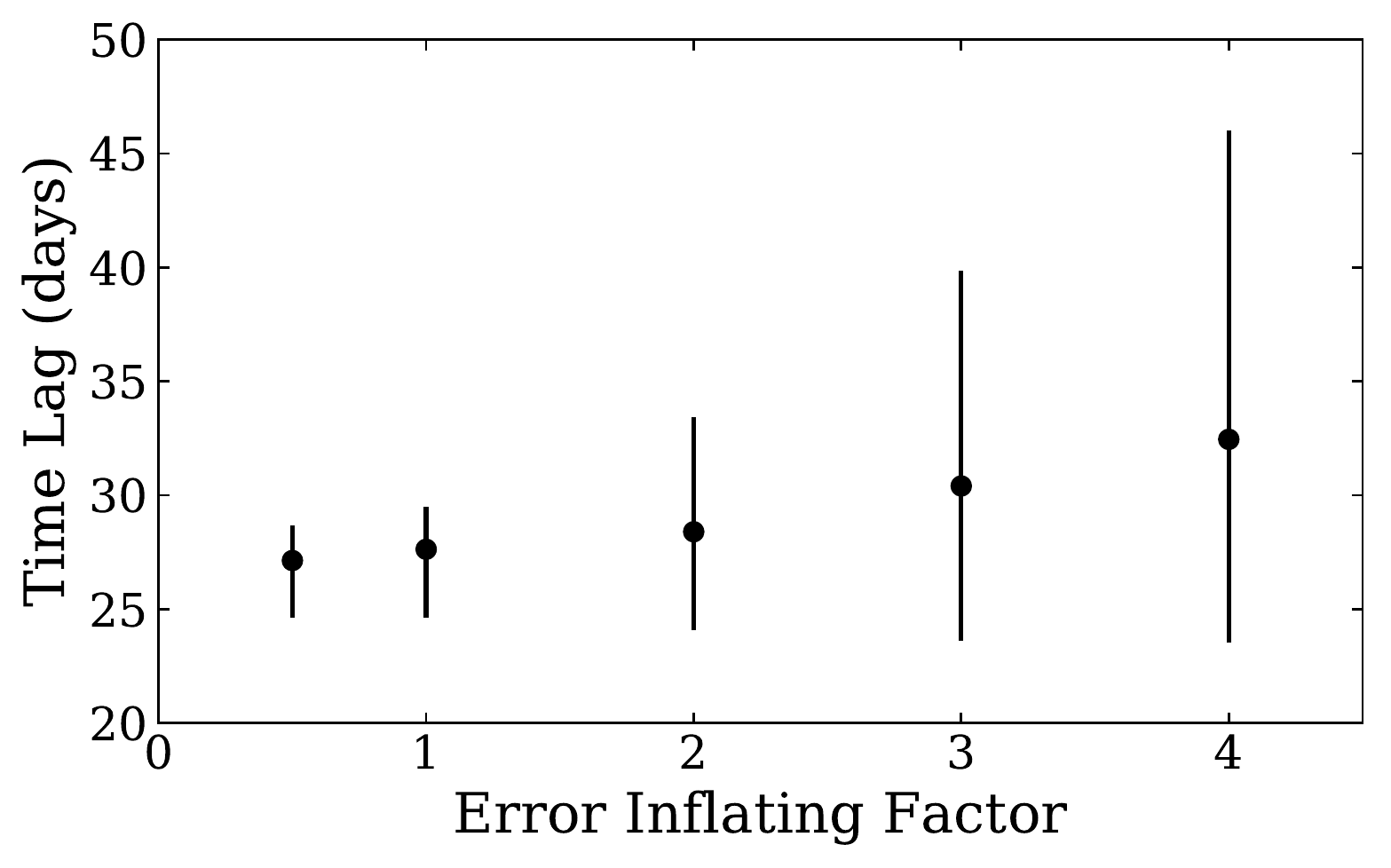}
	\end{tabular}
    \caption{Effects of the SNR on estimates of the composite lag. The top panel shows the coadded ZDCFs from the mock sample generated from the \citet{Bentz_etal_2013} local RM sample (see \S\ref{sec:localRM}) as a function of the constant error inflating factor applied to the sample. The gray shaded band shows the expected signal from random, uncorrelated light curves for the original SNR. The bottom panel shows the average lag measured from the coadded ZDCF as a function of the error inflating factor. As the SNR decreases, the significance of the coadded ZDCF decreases and the uncertainty in the measured average lag increases, as expected. Nevertheless, there is no significant bias in the measured average lag as SNR decreases, suggesting that the coadding technique is robust even for low SNR data.} 
    \label{fig:sn_test}
\end{figure}

Fig.\ \ref{fig:sn_test} shows the coadded ZDCFs and measured average lags as a function of SNR (i.e., the error inflating factor). As SNR decreases, the significance of the coadded ZDCF drops and the uncertainty\footnote{The uncertainty in the measured average lag includes both the measurement uncertainty from the noise in the LCs and the systematic uncertainty from the sample variance and the coadding technique.} in the measured coadded lag increases, as expected. However, there is no significant bias in the inferred average lags when the SNR decreases, indicating that the coadding technique is stable against SNR. In addition to lowering the SNR of the LCs, we also tested a case where we reduce the LC errors by 50\% and show the results in Fig.\ \ref{fig:sn_test}. As expected, the resulting coadded ZDCF has a stronger signal and the composite lag has a smaller uncertainty than before. This demonstrates the general utility of the coadding technique to strengthen a detection.

{We note that lowering the SNR of the light curves is equivalent to decreasing the intrinsic variability of the light curves while keeping the SNR fixed. Lower-variability light curves require better SNR to be able to detect the correlated signals. On the other hand, the composite method cannot boost the signal-to-noise of the coadded lags indefinitely. If the individual light curves are of too poor quality, then even the composite method will not be able to measure a meaningful average lag of the sample. }

\begin{figure}
\centering
	\begin{tabular}{@{}cc@{}}
	\includegraphics[width=0.48\textwidth]{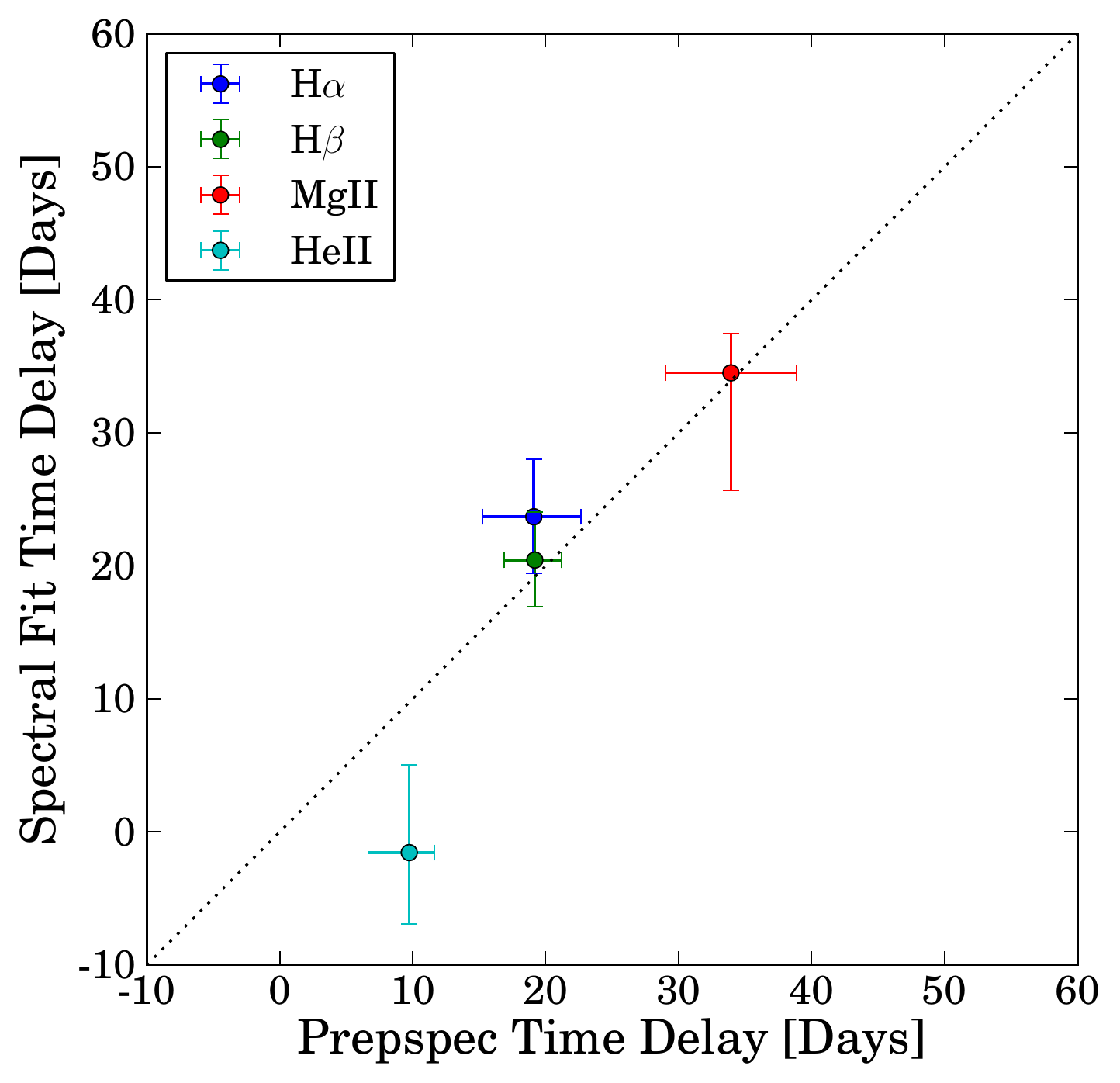}
	\end{tabular}
    \caption{Comparison of the observed composite lags based on light curve fluxes measured with a different spectral fitting approach with the fiducial results based on PrepSpec. Due to the more complicated decomposition in the alternative spectral fits, the flux measurements have larger uncertainties, which led to the non-detection of the \HeII\ lag. }
    \label{fig:flux_test}
\end{figure}

\begin{figure*}
\centering
	\begin{tabular}{@{}cc@{}}
	\includegraphics[width=0.48\textwidth]{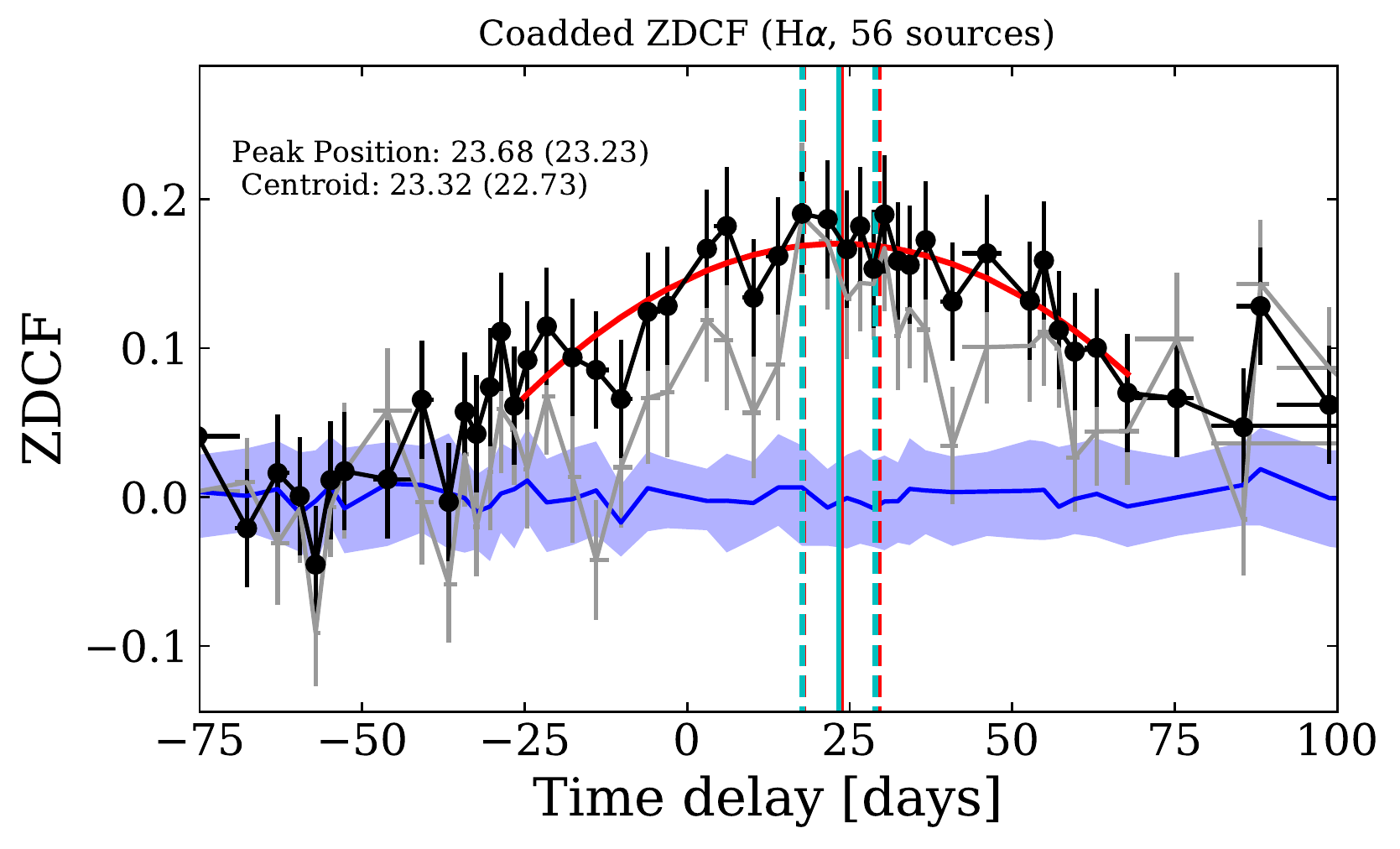}
        \includegraphics[width=0.48\textwidth]{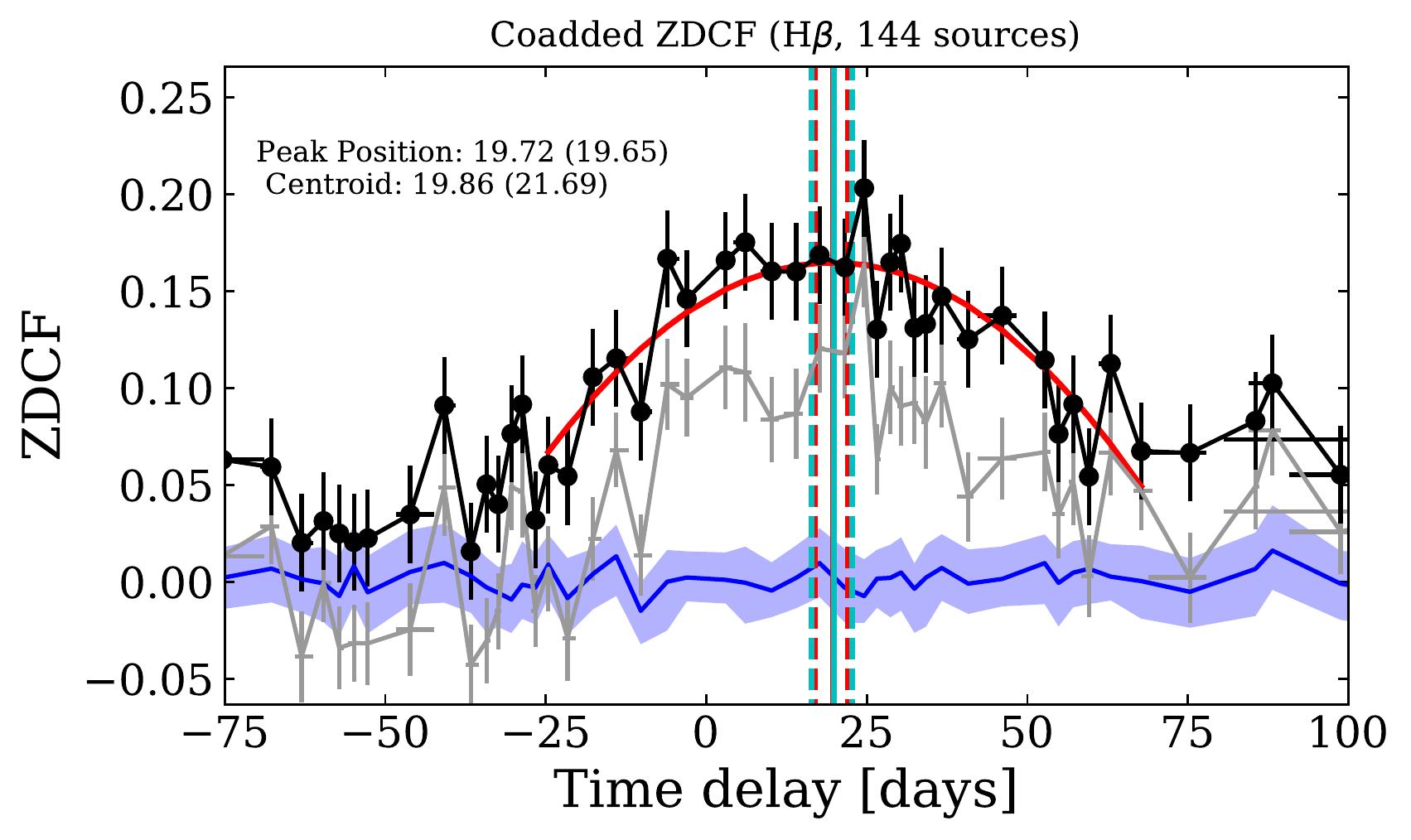} \\
        \includegraphics[width=0.48\textwidth]{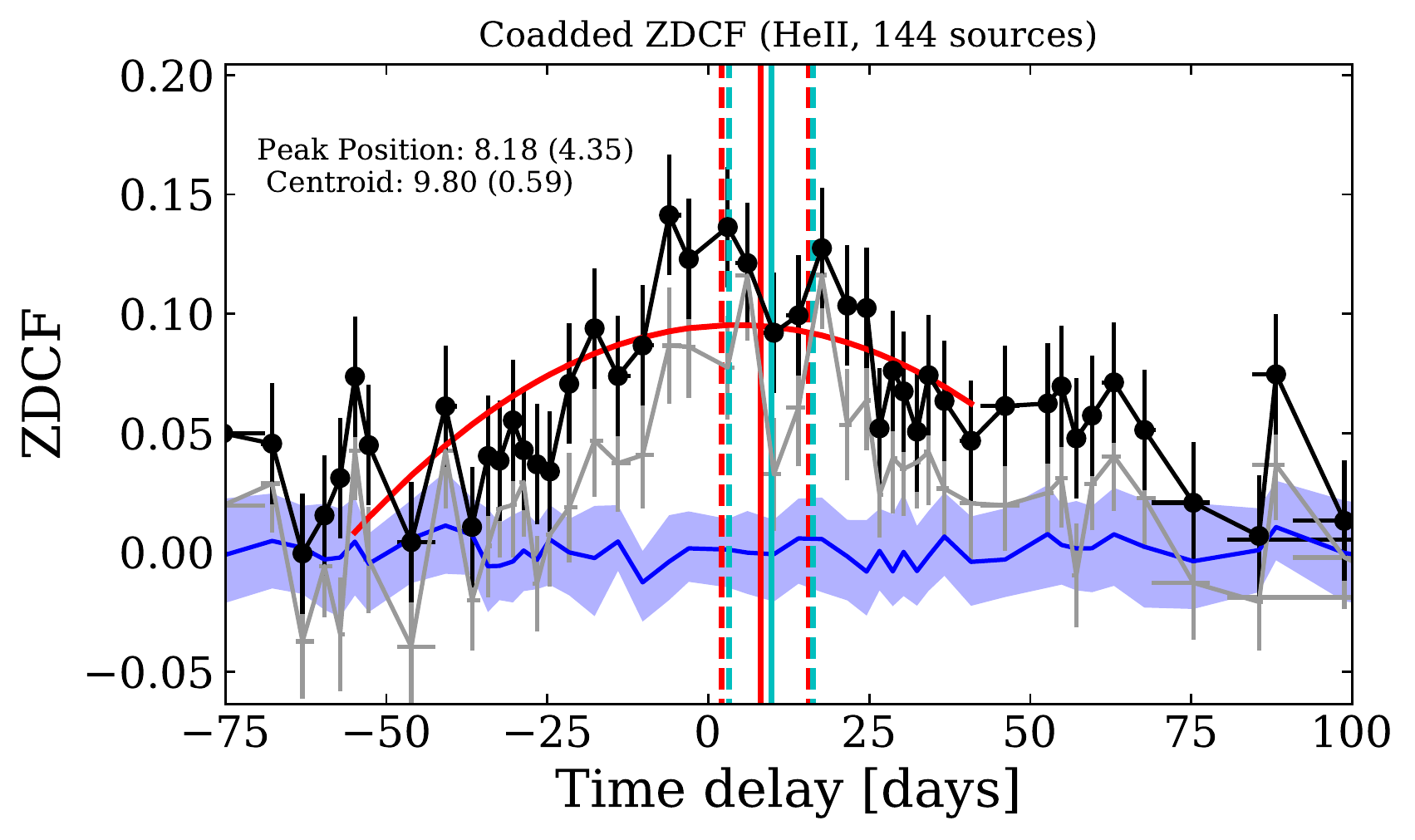}
        \includegraphics[width=0.48\textwidth]{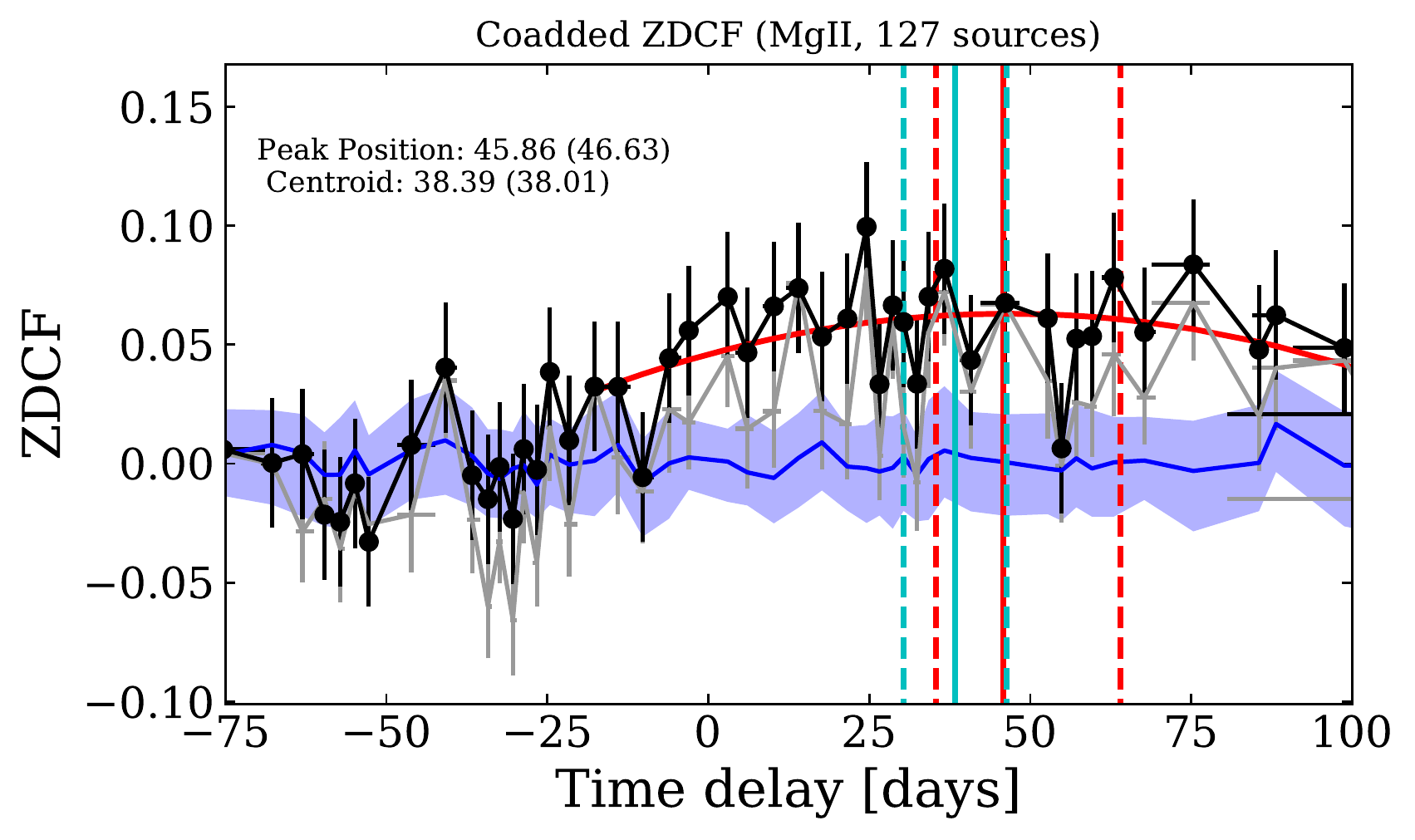}
	\end{tabular}
    \caption{Coadded ZDCFs (points) for the four broad emission lines considered in this work. In each panel, the blue line and the shaded band are the expected signal and uncertainty (16th and 84th percentiles of the distribution) generated from mock light curves with no intrinsic lags. The black data points show the coadded ZDCF with the weighted mean, which we take as the fiducial coadded results. Red curves demonstrate polynomial fits to the coadded ZDCF $\pm$ 50 days around the peak. Bootstrap resampling is used to estimate the uncertainty in the peak measurements. The estimated peak of the correlation (median of the bootstrap distribution) and its $1\sigma$ uncertainties are indicated by the red (polynomial fit) and cyan (direct centroid) vertical lines. The values measured from the median of the bootstrap distribution are marked in the upper-left corner, with the values measured from the original ZDCF indicated in the parentheses. For comparison, the grey line shows the coadded ZDCF with the simple median approach, which has a lower but still statistically significant amplitude for the lag detection. }
    \label{fig:zdcf_all}
\end{figure*}

\begin{figure*}
\centering
	\begin{tabular}{@{}cc@{}}
	\includegraphics[width=0.48\textwidth]{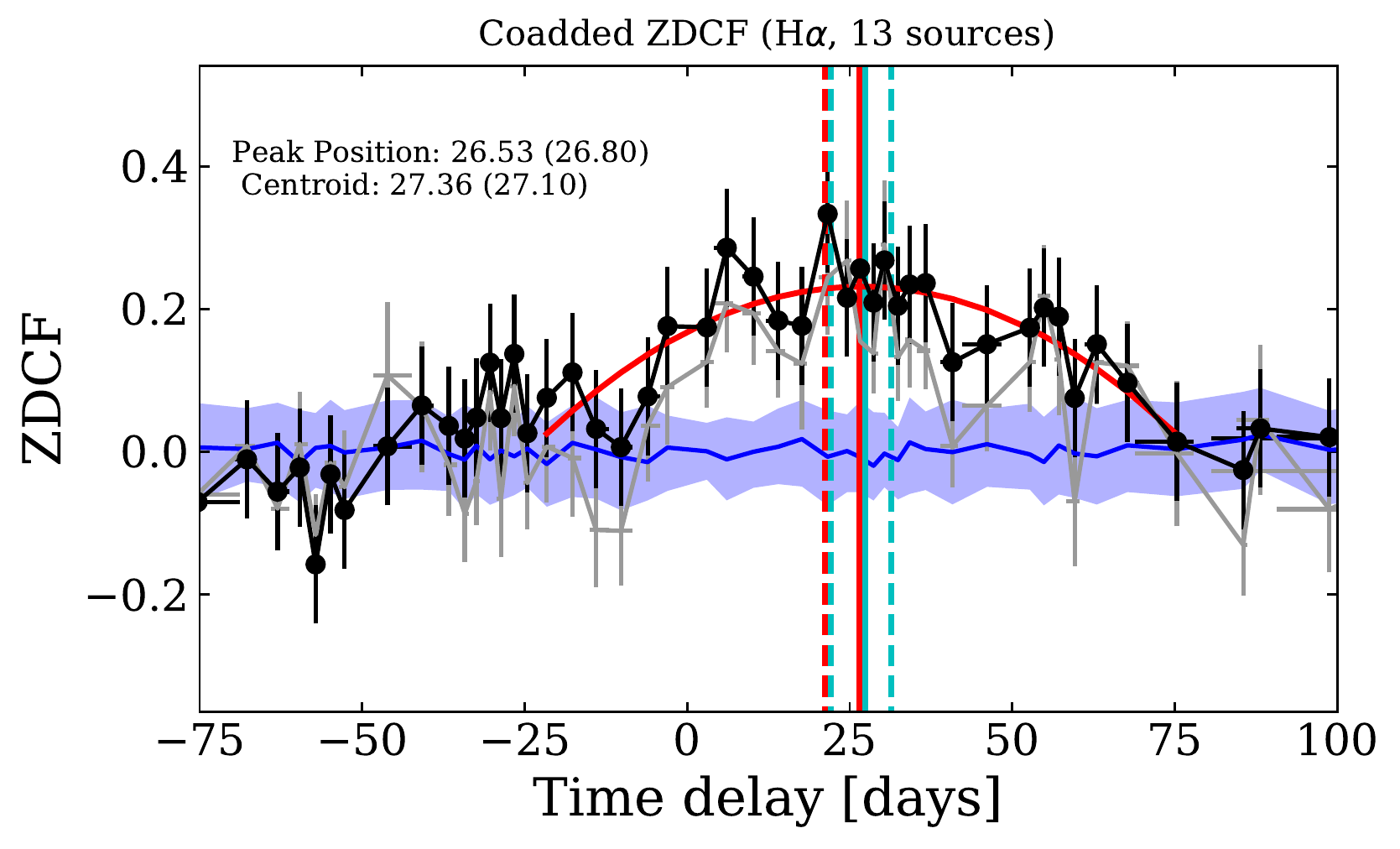}
        \includegraphics[width=0.48\textwidth]{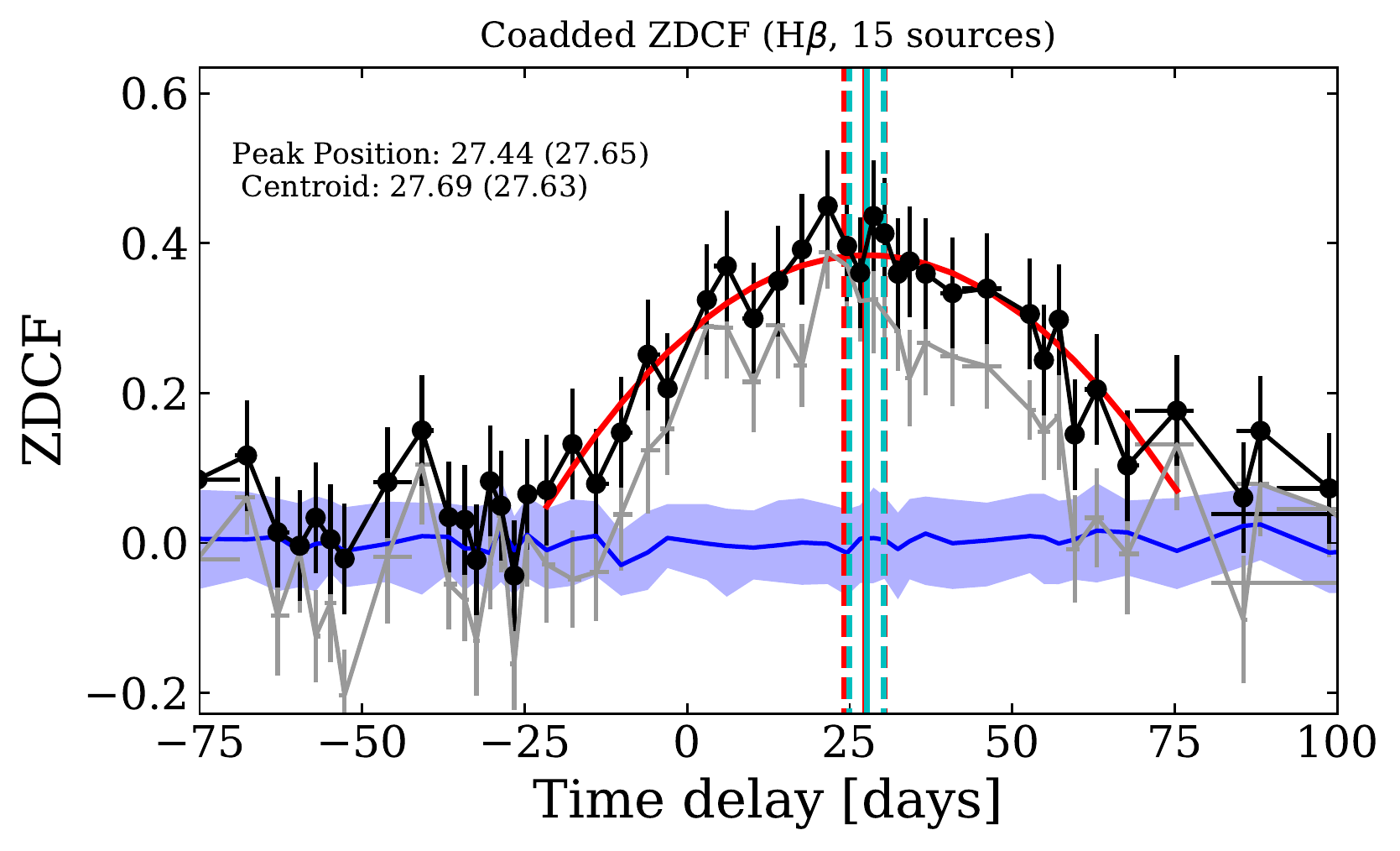} \\
        \includegraphics[width=0.48\textwidth]{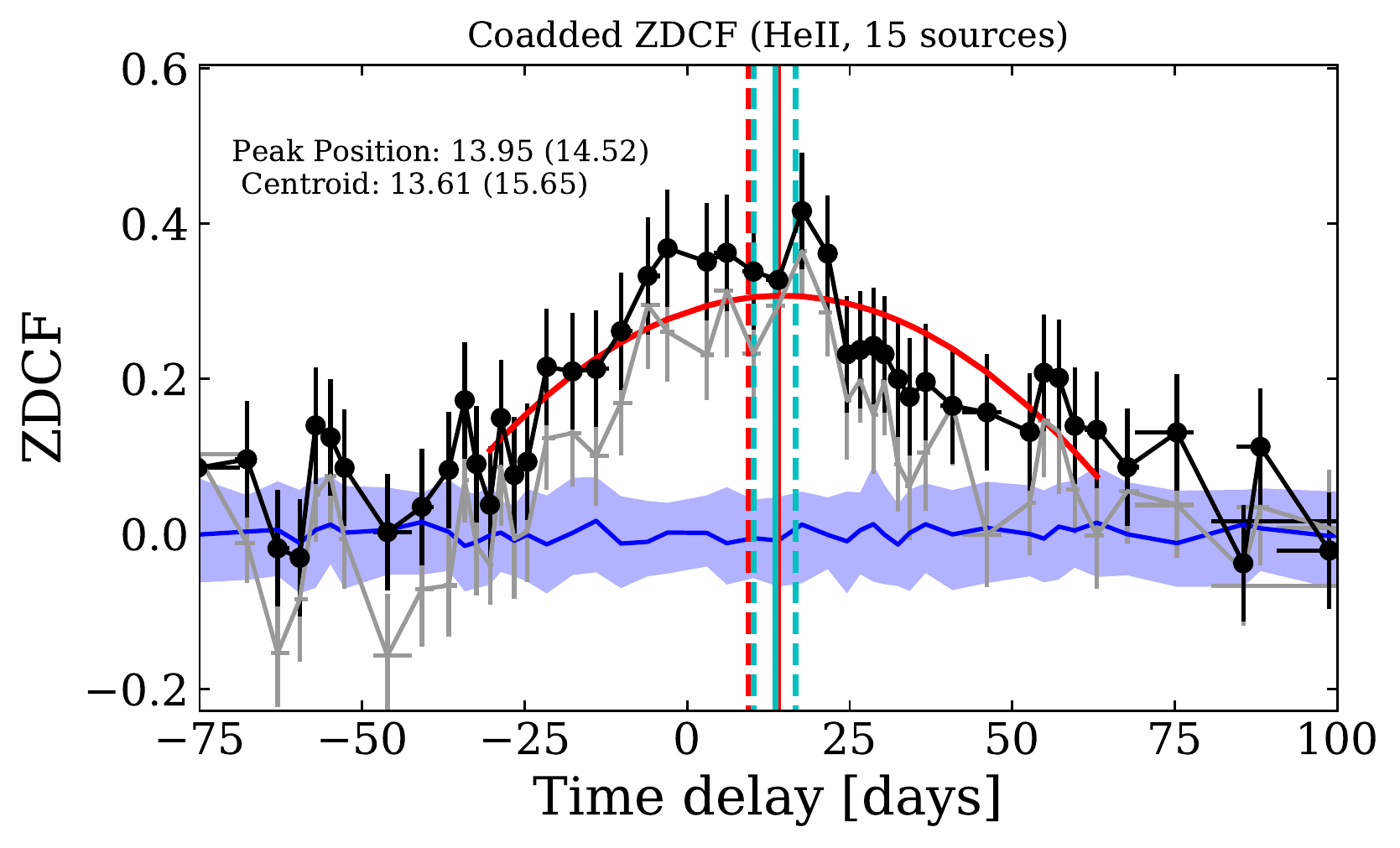}
        \includegraphics[width=0.48\textwidth]{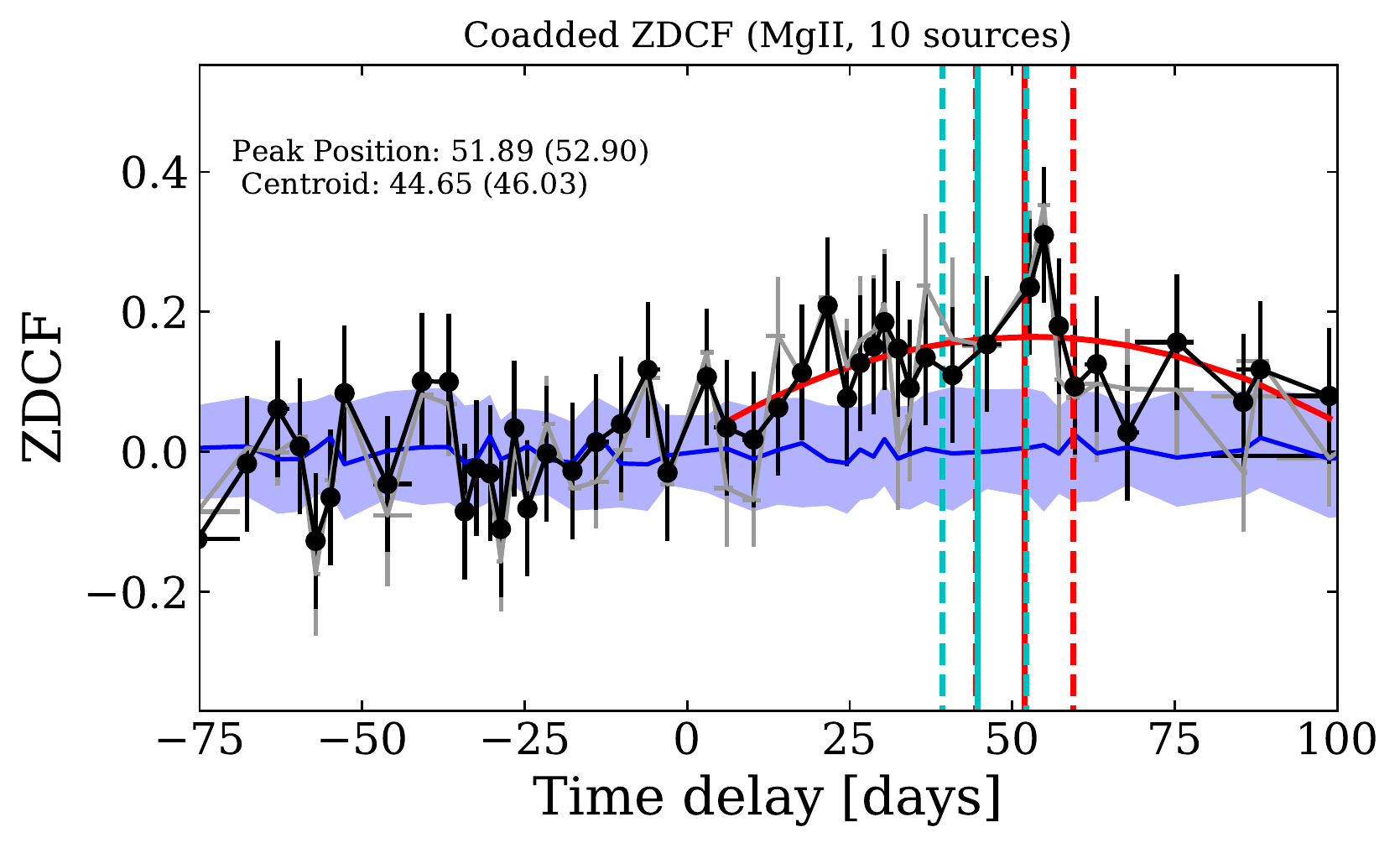}
	\end{tabular}
    \caption{Same format as Figure\ \ref{fig:zdcf_all}, but for the 15 first-lag objects reported in \citet{Shen_etal_2016a}.}
    \label{fig:zdcf_15}
\end{figure*}

\subsubsection{Composite lags based on alternative flux measurements}\label{sec:alter}

Measuring fluxes from a quasar spectra can be a difficult task, which is particularly true for weak broad lines (such as \HeII) and for cases where the decomposition of the continuum, broad- and narrow-line emission is ambiguous. As described in detail in \citet{Shen_etal_2016a}, PrepSpec models each individual spectrum with multiple components to account for the continuum, the broad lines and the narrow lines. It is possible that fluxes measured with alternative approaches differ from the PrepSpec fluxes in a systematic way. Fortunately, the spirit of the composite lag technique is to average out these potential systematic uncertainties in the flux measurements, and provide unbiased average results. Therefore we expect our results are insensitive to systematic uncertainties in the light curve flux measurements in individual objects.

To test the above statement, we use continuum and line light curves measured with the independent spectral fitting approach of \citet[][]{Shen_etal_2011}. A full description of the spectral fits to SDSS-RM quasars will be presented elsewhere (Shen et~al., in preparation). In short, we fit the continuum and \FeII\ emission underneath the broad lines, as well as the adjacent narrow lines, and extract the broad-line flux from functional fits. The main difference between the alternative spectral fits and PrepSpec is the explicit inclusion of the \FeII\ emission in the former approach, but these two approaches also differ in many technical details in how the continuum and lines are modeled. In general, the alternative spectral fits have larger measurement uncertainties in the fluxes due to additional model components (e.g., \FeII\ emission). 

We then apply the same coadding method on the set of light curves based on the alternative flux measurements and measure the average lags. The results for the 15 first-lag objects reported in \citet{Shen_etal_2016a} are compared with the fiducial results based on PrepSpec fluxes, as shown in Figure 6. Coadding all objects yields similar results. We did not detect the \HeII\ lag with the alternative flux measurements (i.e., the lag is consistent with zero) due to larger measurement uncertainties in the light curve fluxes. Nevertheless, this test demonstrates that the average lags measured from the coadding technique are robust against the details of the flux measurements, as expected from the nature of averaging.

\begin{figure}
\centering
    \includegraphics[width=0.48\textwidth]{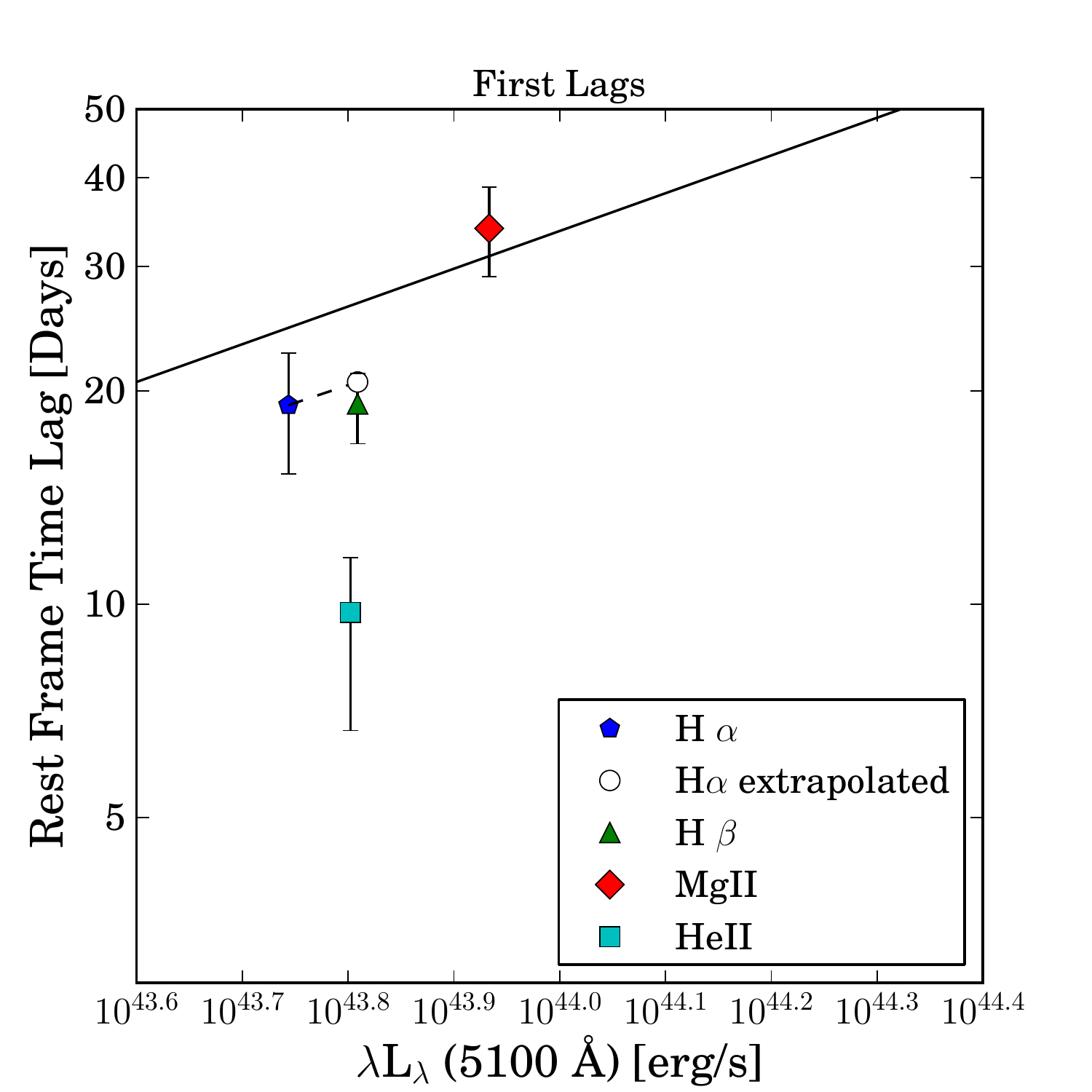} 
    \includegraphics[width=0.48\textwidth]{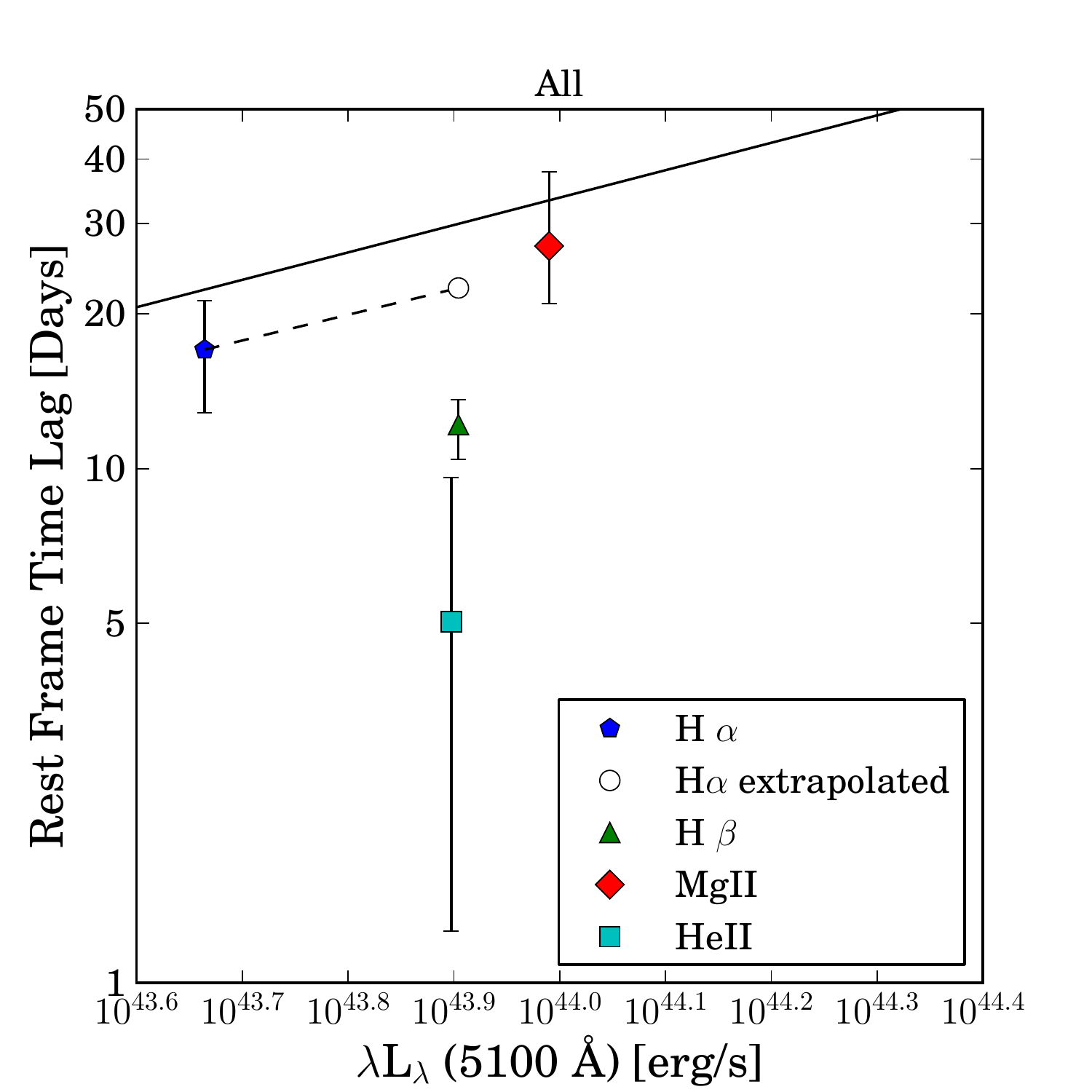}
    \caption{The measured average lags of the four broad emission lines from the polynomial fit to the coadded ZDCF calculated using the weighted mean method, as a function of the weighted mean luminosity of the objects in the coadd (top panel: 15 first-lag objects reported in \citet{Shen_etal_2016a}; bottom panel: all sources in the sample). For \halpha, we also show the expected lag at the same average redshift and luminosity of the \hbeta\ sample as the open symbol (without error bars). The solid line is the measured $R-L$ relation for \hbeta\ in the local RM sample \citep{Bentz_etal_2013}.}
    \label{fig:lag_lum}
\end{figure}

\section{Results}\label{sec:results}

With the extensive tests in \S\ref{sec:tests}, we have demonstrated that the composite lag technique is meaningful, and robust against details in the individual correlation function measurements. We now present the composite lag measurements for our SDSS-RM sample.

\subsection{Coadded lags and errors}\label{sec:result1}

Figure\ \ref{fig:zdcf_all} presents the coadded ZDCFs for the four lines using all objects with available PrepSpec light curves. In all cases, a peak with the line lagging the continuum is present in the ZDCF. However, since the SDSS-RM quasars generally have lower variability amplitudes than those of the local RM AGN sample, the uncertainties in the average lags are substantially larger than in our tests using the local RM sample. 

To demonstrate that the coadded ZDCF is not dominated by a few objects with well-detected individual lags, we remove the 15 objects reported in \citet{Shen_etal_2016a} and calculate the coadded ZDCF for the remaining objects. As these 15 objects were the ones with individual lag detections, they have the ZDCFs with the highest SNR. The results are listed in Table \ref{tab:lag} as the ``other'' samples\footnote{We note that some of the objects in the ``other'' sample can have individually detectable lags if the light curves are of better quality. For example, our recent SDSS-RM work incorporating additional photometric light curves reported additional \halpha\ and \hbeta\ lags (Grier et al.\ 2017). However, these lags are non-detections with the spectroscopic-only light curves used here, and the composite method is the only way to recover an average lag for these objects. }. We still detect statistically significant lags that are consistent with the full sample, demonstrating the effectiveness of the coadding technique. 

As a sanity check, coadding the 15 objects reported in \citet{Shen_etal_2016a}
produces much more significant signals (higher ZDCF peak amplitudes and
narrower peaks) given the better quality of individual ZDCFs for these objects
than for the rest of our sample (see Figure\ \ref{fig:zdcf_15}). The composite \hbeta\ and \MgII\ lags are also consistent with the median of the individual lags reported in \citet{Shen_etal_2016a}. The relative durations of the composite lags for different lines are consistent with results based on the full sample, as discussed below.  

Using the ZDCF and coadding in the observed frame avoids interpolation and rebinning of the individual correlation functions, which will lead to correlated errors and complicate the interpretation. However, a caveat of this approach is that objects at different redshifts will have different time dilations of their intrinsic lags. This will result in an additional broadening of the coadded correlation function compared to stacking in the rest-frame of these quasars. We have tested coadding the individual ZDCFs in the rest frame of the quasars by shifting and interpolating the original ZDCFs onto a common rest-frame time-lag grid. We found consistent results (to within $\sim 1\sigma$) and uncertainties compared to the stacks performed in the observed frame after scaling the latter by a factor of $1+\bracket{z}$. The results for the rest-frame stacks are provided in the last three columns in Table 1 for completeness. Stacking in the rest frame does not produce significantly smaller error bars for the average lags, which may be due to the possibility that objects with different luminosities (which presumably have different lags) already significantly broaden the stacked ZDCF, and the additional broadening due to redshfit is sub-dominant. If we assume the $R\propto L^{0.5}$ relation from \citet{Bentz_etal_2013}, the dispersion in the observed-frame lags due to luminosity is a factor of $\sim 2-3$ larger than that due to the $(1+z)$ factor in the four line samples.  

\subsection{Lags for different line species}

Figure\ \ref{fig:lag_lum} shows the average lags as a function of sample-averaged continuum luminosity for the four lines, where the lags have been shifted to the rest frame using the sample-averaged redshift. To remove host-starlight contamination from the rest-frame 5100\,\AA\ continuum-luminosity measurements, we adopted a spectral-decomposition technique detailed in \citet{Shen_etal_2015b} to derive the quasar-only luminosity. We use the same weights as for the coadded ZDCFs to compute a weighted-mean luminosity for the sample. The \halpha\ sample has a relatively lower average redshift than other three line samples, and thus has the lowest average luminosity. For the other three lines, their samples have similar continuum luminosities and redshifts, allowing a fair comparison of their average lags. {If we extrapolate the average \halpha\ lag to higher continuum luminosity assuming $\tau\propto L^{0.5}$ \citep[i.e., consistent with the measured $R-L$ relation for \hbeta,][]{Bentz_etal_2013}, we derive the open data point in Figure\ \ref{fig:lag_lum} at the sample averaged luminosity of the \hbeta\ sample.} Although the error bars overlap, Figure\ \ref{fig:lag_lum} reveals some evidence that \HeII\ has, on average, shorter lags than \hbeta, and \hbeta\ has shorter average lags than \halpha. While having large error bars, the composite \MgII\ lag appears to be slightly longer than that for \halpha. These results are qualitatively consistent with earlier individual RM measurements of different line species in the same objects \citep[e.g.,][]{Peterson_Wandel_1999,Peterson_etal_2004,Bentz_etal_2010,Grier_etal_2012b,Grier_etal_2013}. In addition, the tentative evidence that \MgII\ has on average longer lags than \hbeta\ is also consistent with the finding that \MgII\ varies less than \hbeta\ on the same timescales \citep[e.g.,][]{Sun_etal_2015}, suggesting that the \MgII\ gas may be located slightly further out than the \hbeta\ gas.

However, Figure\ \ref{fig:lag_lum} suggests that the average \hbeta\ lag is significantly shorter than the expected lag from the measured $R-L$ relation based on individual RM measurements in the low-$z$ sample \citep[e.g.,][]{Bentz_etal_2013}. We speculate that the main reason for this discrepancy is due to imperfectly weighted mean redshifts and luminosities for the sample.  When averaging the ZDCF for a sample covering a wide redshift and luminosity range, objects that have more obvious (higher amplitude) correlations contribute more power to the peak location in the final coadded ZDCF in a non-trivial way. Lower-$z$ and lower-luminosity objects are likely to have a stronger influence on the determination of the composite lag than higher-$z$ and higher-luminosity objects, given the higher S/N in the light curves and stronger intrinsic variability. As a result, the sample-averaged redshift and luminosity using the weights based on measurement errors in the individual ZDCFs (instead of the amplitude of the ZDCFs) may overestimate the true sample-averaged redshift and luminosity. Unfortunately, without a robust weighting scheme to account for this complication, we can only qualitatively explain this discrepancy, but cannot correct the bias quantitatively. The solution would be to coadd objects roughly at the same redshift and luminosity, which should become possible when we have better individual ZDCF measurements (see \S\ref{sec:sum}) or with future, larger MOS-RM samples. Nevertheless, this caveat should affect all lines for the same set of objects, and hence the relative lags of different lines should be robust. 

Another possible explanation of the discrepancy is that most of the objects included in the coadded ZDCF have different accretion rates and hence different spectral energy distributions than those for the local sample used to measure the $R-L$ relation for \hbeta. Recently, \citet{Du_etal_2015,Du_etal_2016} suggested that objects with higher accretion rates have significantly shorter BLR lags compared to lower accretion-rate objects. Quasars with higher Eddington ratios also vary less than those with lower Eddington ratios \citep[e.g.,][]{Ai_etal_2013}, leading to a potential selection bias in the sample with robust lag detections or target samples for RM campaigns. The local sample used to measure the $R-L$ relation is dominated by relatively low accretion-rate objects \citep[e.g., Figure \ 1 of][]{Shen_etal_2015a} compared to the general population of SDSS-RM quasars. Although it is unlikely that this is the complete explanation, we will investigate this possibility further in future work to determine if it can account for at least some of the discrepancy seen here. 

It is worth noting that our recent work on individual lag measurements based on both spectroscopic and photometric data (Grier et~al.\ 2017) measured $\sim 40$ \hbeta\ lags and $18$ \halpha\ in the SDSS-RM sample. Complementary to the composite approach, these individual \hbeta\ lags are also significantly shorter than the expectation from the local $R-L$ relation on average. We have tested by coadding the objects with detections in Grier et~al.\ (2017) using spectro-only LCs. We got restframe composite lags of $16_{-4.6}^{+5.3}$ days for \halpha\ and $14_{-1.8}^{+1.8}$ days for \hbeta, fully consistent with the average lags from Grier et~al. (2017). This again demonstrates that our composite approach is robust in recovering a sample-averaged lag.

\begin{figure*}
\centering
	\begin{tabular}{@{}cccc@{}}
	\includegraphics[width=0.48\textwidth]{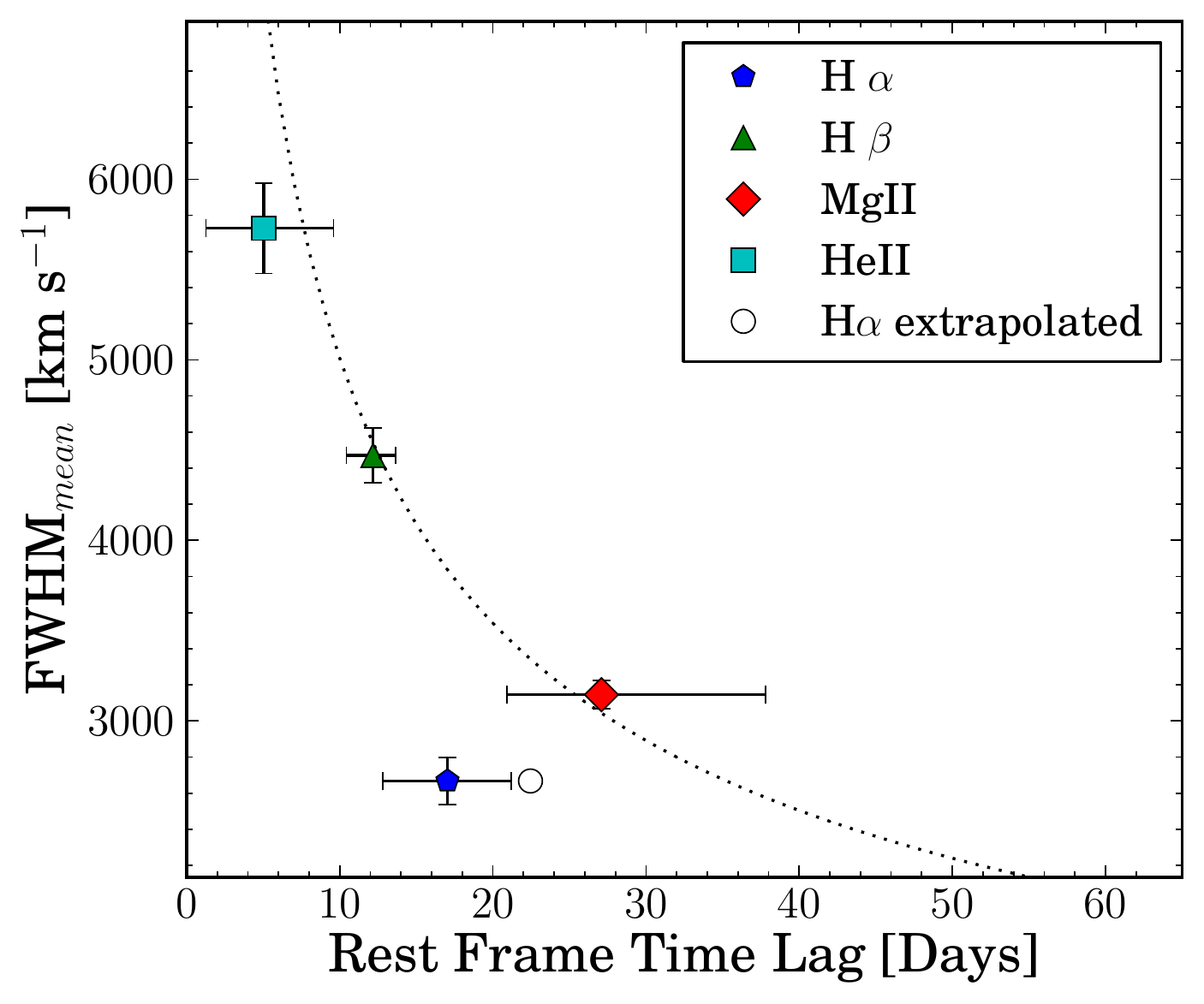}
        \includegraphics[width=0.48\textwidth]{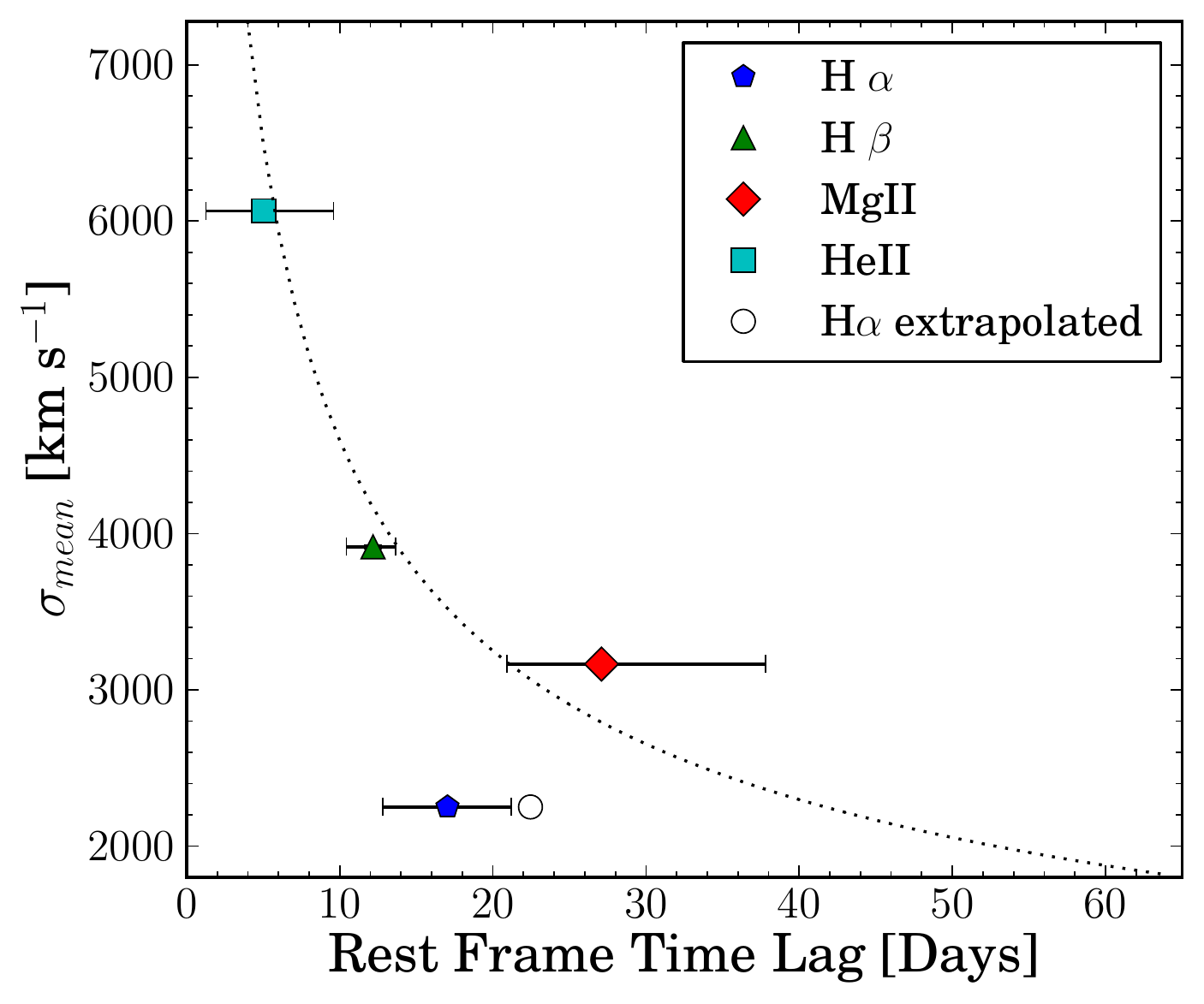} \\
        \includegraphics[width=0.48\textwidth]{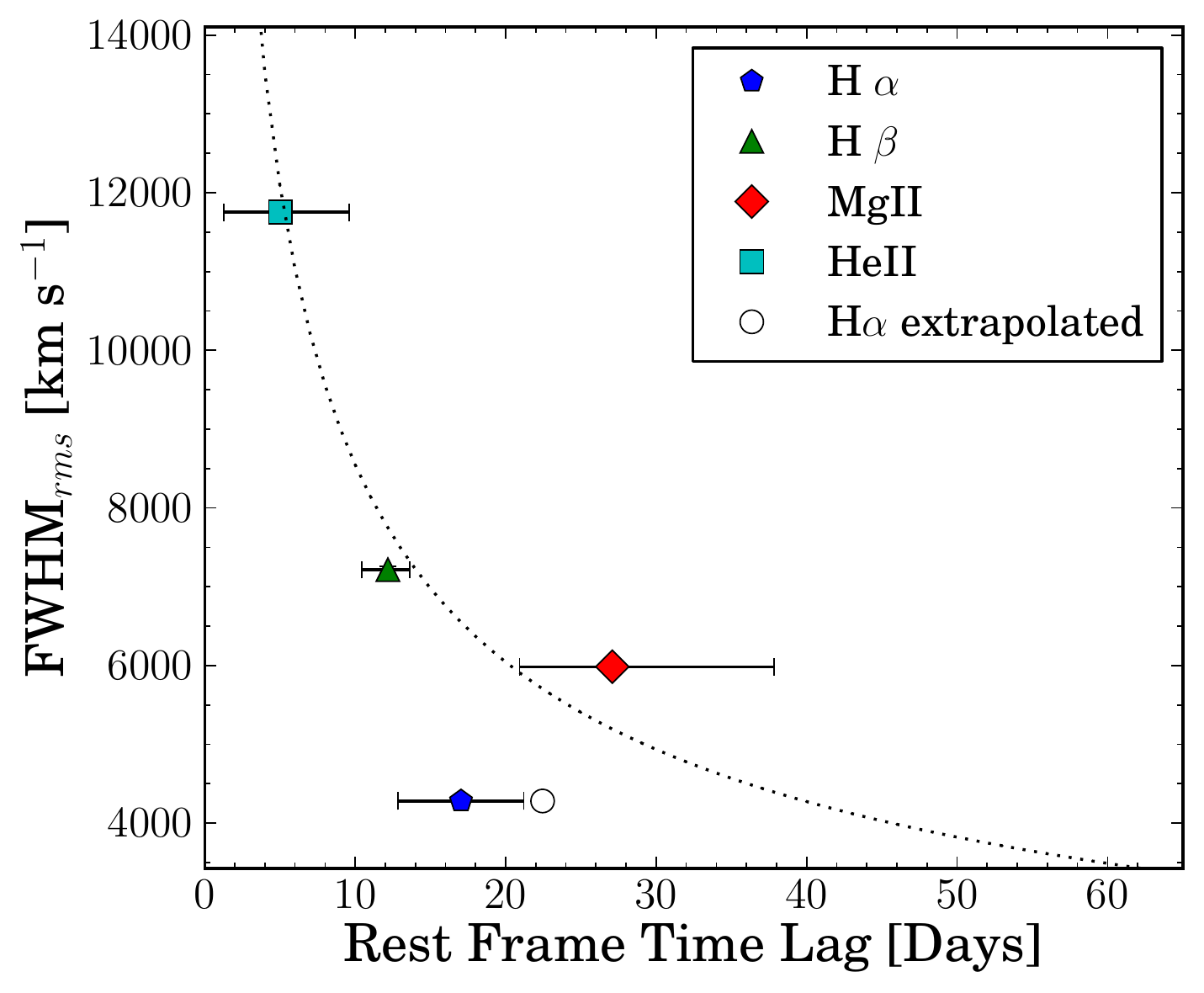}
        \includegraphics[width=0.48\textwidth]{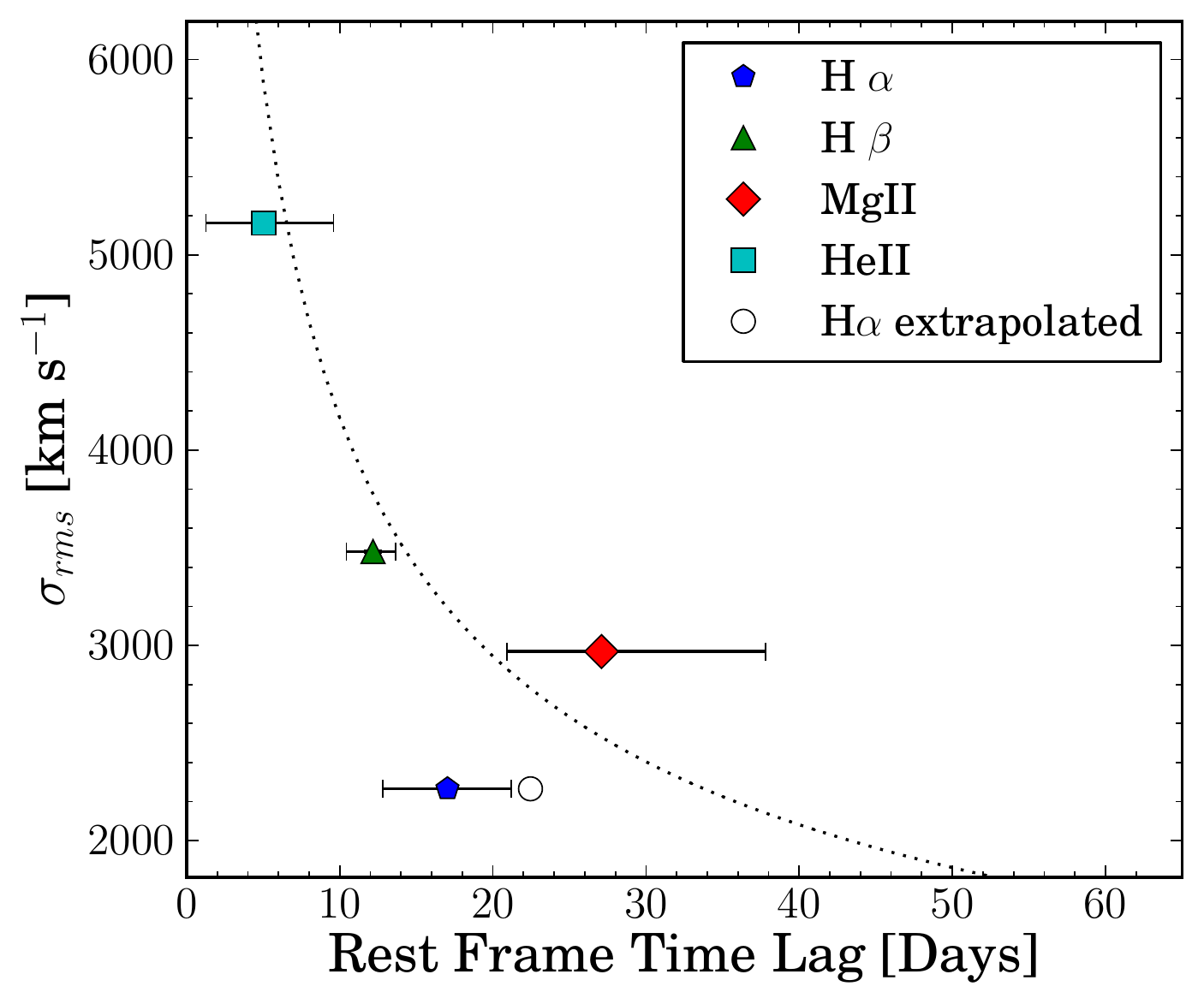}
	\end{tabular}
    \caption{Relations between the average lag (from the polynomial fit) and line width for the four lines. The average lags have been shifted to rest frame using the sample-averaged redshifts in Table 1. The adopted line widths are placed at the weighted mean line width for each sample, the uncertainty of which is estimated using propagated individual measurement uncertainties, and is almost negligible. We use four definitions of the line width. In each panel, the dotted line indicates the virial relation $V\propto \tau^{-0.5}$, fit to all data points except for that of \halpha, for which the sample average luminosity is quite different from those of the other three lines. The open symbol in each panel is the expected lag (without error bars) for \halpha\ at the same average redshift and luminosity of the \hbeta\ sample. }
    \label{fig:lag_width}
\end{figure*}

\subsection{Luminosity and line-width dependence of time lags}

We further divide each sample at the median luminosity and investigate the luminosity dependence of the average lags. For \halpha,
a clear luminosity dependence of the average lag is seen (time lags are shorter for the lower luminosity subsample) after the $(1+\bracket{z})$ time dilation is included (see Table 1), and is roughly consistent with the measured $R-L$ relation for \hbeta\ \citep[e.g.,][]{Bentz_etal_2013}. However, the luminosity trend is poorly constrained for \hbeta, \HeII\ and \MgII, both due to the weak coadded ZDCF signals in the divided samples, and to the narrow dynamic range in luminosity covered by the full samples. We further tested dividing the sample according to the expected lags in the observed frame using the $R-L$ relation for \hbeta\ in the local sample \citep{Bentz_etal_2013}, but the resulting composite lags are again too noisy to reveal any conclusive trends for \hbeta, \HeII\ and \MgII. We plan to investigate the luminosity dependence further with the inclusion of photometric data in our analysis in the future. 



Figure \ \ref{fig:lag_width} shows the relationship between the average lag
and the sample-averaged (weighted-mean) line width for the four lines. We use
four definitions of line width: the line dispersion (square root of the second moment of the line) $\sigma_{\rm line}$ measured from the mean and rms spectra, and the full-width-at-half-maximum (FWHM) measured from the mean and rms spectra. The rms spectra are produced by PrepSpec after decomposing the line emission and the continuum emission. All four line-width definitions are consistent with the virial prediction, and the luminosity-extrapolated \halpha\ point matches as well. Lines with shorter lags in general have larger velocity widths \citep[e.g.,][]{Peterson_Wandel_1999,Grier_etal_2013}, consistent with the scenario that there is a stratification of the BLR for different lines with different ionization potentials such that high-ionization lines (e.g., \HeII) are on average closer to the BH and hence have larger line widths. 

\section{Conclusions}\label{sec:sum}

Using the large sample of quasars with RM data from the SDSS-RM project, we have tested the feasibility of measuring composite lags for statistical samples. We applied the ZDCF method on the $z\lesssim 1$ subset of the SDSS-RM quasar sample, and measured composite lags for \halpha, \hbeta, \HeII\ and \MgII\ at average redshifts $>0.3$. Compared to the earlier work on composite RM \citep[][]{Fine_etal_2013}, our work focused on a different regime of redshift and luminosity with the unique SDSS-RM sample, and provided composite-lag measurements for broad lines other than \MgII\ and \CIV. The findings from this work are the following:

\begin{enumerate}

\item[1.] When luminosity and redshift are matched, the sample-averaged lag decreases in the order of \MgII, \halpha, \hbeta, and \HeII, suggesting that there is a stratification of the BLR, with high-ionization lines closer to the ionizing continuum than low-ionization lines. In particular, \MgII\ may have slightly longer lags on average than \hbeta, suggesting that the \MgII\ gas may be located at a larger distance from the black hole than the \hbeta\ gas. 

\item[2.] Lines with shorter average lags have larger average line widths. The relation between average line width and lag is roughly consistent with the virial relation. 

\end{enumerate}

These results are in qualitative agreement with earlier RM studies based on individual objects and at lower redshifts \citep[e.g.,][]{Peterson_Wandel_1999,Peterson_etal_2004, Bentz_etal_2010,Grier_etal_2012b,Grier_etal_2013}, and are among the first results on stratified BLR structure at $z>0.3$ \citep[e.g.,][]{Fine_etal_2013}, particularly for \HeII\ lags.

However, as discussed in \S\ref{sec:results}, it is challenging to assign an average redshift and luminosity to the sample, due to the non-trivial contributions from individual objects to the coadded ZDCF. For flux-limited samples that cover broad ranges of redshift and luminosity, the simple median or weighted mean redshift and luminosity may be overestimated. This caveat can be avoided by focusing on samples within a narrow redshift-luminosity range, at the cost of significantly degrading the sample statistics. Finally, as pointed out in earlier work \citep[][]{Fine_etal_2013,Brewer_Elliott_2014}, it is somewhat ambiguous to interpret the stacked correlation function for the underlying sample other than the indication of a typical ``average'' lag from the peak of the stacked correlation. It is possible to deploy more sophisticated statistical inferences to extract information about the intrinsic-lag distribution of the underlying sample from the stacked analysis \citep[e.g.,][]{Brewer_Elliott_2014}.

SDSS-RM continues to monitor the same quasar sample to extend the time baseline for the detection of long lags at higher redshifts. In addition, we are incorporating the more densely sampled photometric light curves into our time-series analyses. These photometric light curves will improve the spectrophotometry of our spectroscopic epochs with overlapping photometric epochs. With the addition of photometric light curves to enhance the correlation signals in individual objects and the extended time baseline, we plan to expand the redshift coverage to $z>1$, to include additional broad lines (such as \CIII\ and \CIV), as well as to improve greatly the quality of composite lag measurements in subsamples divided by luminosity and other quasar parameters.  

\acknowledgements

We thank Brad Peterson for useful comments on the draft, and the anonymous referee for suggestions that significantly improved this work. 
JIL and YS acknowledge support from an Alfred P. Sloan Research Fellowship and NSF grant AST-1715579. WNB, CJG, DPS and JRT acknowledges support from NSF grant AST-1517113. KH acknowledges support from STFC grant ST/M001296/1. LCH was supported by the National Key Program for Science and Technology Research and Development (2016YFA0400702) and the National Science Foundation of China (11303008, 11473002).

Funding for SDSS-III has been provided by the Alfred P. Sloan Foundation, the
Participating Institutions, the National Science Foundation, and the U.S.
Department of Energy Office of Science. The SDSS-III web site is
http://www.sdss3.org/.

SDSS-III is managed by the Astrophysical Research Consortium for the
Participating Institutions of the SDSS-III Collaboration including the
University of Arizona, the Brazilian Participation Group, Brookhaven National
Laboratory, University of Cambridge, Carnegie Mellon University, University
of Florida, the French Participation Group, the German Participation Group,
Harvard University, the Instituto de Astrofisica de Canarias, the Michigan
State/Notre Dame/JINA Participation Group, Johns Hopkins University, Lawrence
Berkeley National Laboratory, Max Planck Institute for Astrophysics, Max
Planck Institute for Extraterrestrial Physics, New Mexico State University,
New York University, Ohio State University, Pennsylvania State University,
University of Portsmouth, Princeton University, the Spanish Participation
Group, University of Tokyo, University of Utah, Vanderbilt University,
University of Virginia, University of Washington, and Yale University.


\end{document}

%% file: table_lag_efix.tex
\begin{table*}
\caption{Composite Lags}\label{tab:lag}
\centering
\scalebox{0.72}{
\begin{tabular}{crrr|rrrrrr|rrrrrr|rrrr|rrr}
\hline\hline
Line & $N_{\rm obj}$ & $\bracket{z}$ & $\bracket{\log L_{\rm 5100}}$ & \multicolumn{6}{|c}{$\bracket{\tau}$ (Simple Median)} & \multicolumn{6}{|c}{$\bracket{\tau}$ (Weighted Mean)} & \multicolumn{4}{|c}{$\bracket{\textrm{line width}}$} & \multicolumn{3}{|c}{$\bracket{\tau_{\rm rest}}$ (Rest Frame)} \\

& & & $[{\rm erg\,s^{-1}}]$ & \multicolumn{3}{c}{polynomial fit} &  \multicolumn{3}{c}{direct centroid} & \multicolumn{3}{|c}{\bf polynomial fit} &  \multicolumn{3}{c}{direct centroid} & \multicolumn{4}{|c}{[$\kms$]} & \multicolumn{3}{|c}{polynomial fit}  \\

& & & & median & $16\%$ & $84\%$ & median & $16\%$ & $84\%$ & {\bf median} & {\bf 16\%} & {\bf 84\%} & median & $16\%$ & $84\%$ & $\sigma_{\rm mean}$ & $\sigma_{\rm rms}$ & FWHM$_{\rm mean}$ & FWHM$_{\rm rms}$ &  median & $16\%$ & $84\%$ \\

\hline
\halpha\ all & 56  & 0.39 & 43.66 & 26.4 & 17.7 & 35.0 & 27.9 & 19.1 & 35.1 & 23.7 & 17.8 & 29.5 & 23.3 & 17.7 & 29.0 & 2251 & 2266 & 2668 & 4281  & 19.7 & 15.3 & 24.6 \\
other           & 43  & 0.39 & 43.64 & 22.9 & 13.8 & 41.0 & 24.3 & 14.4 & 38.9 & 21.7 & 13.2 & 34.4 & 21.8 & 13.5 & 31.4 & 2473 & 2479 & 2681 & 4662  & 19.2 & 11.9 & 28.2 \\
first lags      & 13  & 0.39 & 43.74 & 29.0 & 17.4 & 37.0 & 30.6 & 20.5 & 38.9 & 26.5 & 21.2 & 31.4 & 27.4 & 22.1 & 31.3 & 1464 & 1511 & 2621 & 2932  & 19.5 & 16.0 & 22.7 \\
low-$L$       & 28  & 0.32 & 43.29 & 17.3 & 6.5  & 27.1 & 17.0 & 8.5  & 25.4 & 16.1 & 7.1  & 23.3 & 15.4 & 6.5  & 23.3 & 2306 & 2284 & 2842 & 4424  & 14.9 & 8.6  & 20.0 \\
high-$L$     & 28  & 0.45 & 44.02 & 52.1 & 33.0 & 63.7 & 47.6 & 34.8 & 57.6 & 41.6 & 31.9 & 58.8 & 38.2 & 30.1 & 48.5 & 2196 & 2247 & 2493 & 4138  & 27.3 & 19.1 & 36.0 \\

\hline

\hbeta\ all  & 144 & 0.62 & 43.90 & 22.7 & 18.4 & 26.2 & 24.2 & 18.7 & 26.9 & 19.7 & 16.9 & 22.1 & 19.9 & 16.4 & 22.7 & 3915 & 3481 & 4471 & 7216  & 12.7 & 10.6 & 15.2 \\
other                     & 129 & 0.64 & 43.92 & 18.4 & 11.3 & 24.5 & 19.1 & 12.4 & 25.5 & 15.9 & 11.7 & 19.5 & 15.0 & 11.0 & 19.0 & 4153 & 3696 & 4553 & 7628  & 9.3  & 6.5  & 11.9 \\
first lags                & 15  & 0.43 & 43.81 & 26.4 & 23.8 & 29.8 & 27.8 & 25.5 & 31.3 & 27.4 & 24.1 & 30.3 & 27.7 & 24.9 & 30.2 & 2107 & 1842 & 3839 & 4076  & 20.0 & 18.2 & 22.1 \\
low-$L$                   & 72  & 0.51 & 43.59 & 21.5 & 17.1 & 24.7 & 21.7 & 16.0 & 25.1 & 19.0 & 16.6 & 21.6 & 18.3 & 15.1 & 21.8 & 3680 & 3363 & 4707 & 6983  & 12.4 & 9.9  & 14.1 \\
high-$L$                  & 72  & 0.73 & 44.21 & 24.9 & 11.7 & 44.4 & 28.0 & 17.5 & 41.4 & 24.6 & 12.1 & 52.9 & 26.4 & 15.5 & 48.3 & 4148 & 3597 & 4237 & 7448  & 14.5 & 9.1  & 36.0 \\

\hline

\HeII\ all   & 144 & 0.62 & 43.90 & 12.7 & 6.4  & 18.3 & 12.4 & 5.3  & 18.7 & 8.2  & 2.0  & 15.6 & 9.8  & 3.2  & 16.3 & 6064 & 5163 & 5728 & 11753 & 3.8  & 1.3  & 7.2  \\
other                     & 129 & 0.65 & 43.91 & 12.7 & 3.2  & 22.0 & 12.7 & 2.5  & 22.7 & 4.5  & -2.2 & 16.0 & 6.1  & -0.8 & 18.0 & 6411 & 5432 & 6006 & 12313 & 3.4  & -0.2 & 8.0  \\
first lags                & 15  & 0.43 & 43.80 & 13.8 & 8.2  & 18.3 & 14.2 & 10.0 & 18.9 & 13.9 & 9.5  & 16.7 & 13.6 & 10.2 & 16.7 & 3374 & 3067 & 3570 & 7402  & 6.5  & 3.5  & 8.9  \\

\hline

\MgII\  all  & 127 & 0.69 & 43.99 & 38.2 & 24.6 & 56.8 & 39.0 & 28.3 & 49.1 & 45.9	& 35.4 & 64.0 & 38.4 & 30.4 & 46.3 & 3166 & 2969 & 3145 & 5985  & 35.9 & 20.9 & 42.9 \\
other                     & 117 & 0.71 & 43.99 & 34.2 & 21.8 & 52.7 & 36.9 & 25.4 & 49.6 & 30.3 & 16.5 & 49.2 & 30.9 & 19.8 & 43.7 & 3294 & 3086 & 3198 & 6209  & 28.9 & 16.3 & 41.7 \\
first lags                & 10  & 0.53 & 43.93 & 52.7 & 40.8 & 58.8 & 43.2 & 30.4 & 52.8 & 51.9 & 44.4 & 59.4 & 44.7 & 39.3 & 52.1 & 1695 & 1621 & 2544 & 3421  & 33.6 & 23.9 & 38.0 \\

\hline
\hline\\
\end{tabular}}
\begin{tablenotes}
      \small
      \item NOTE. --- All lag measurements are in units of days in the observed frame except for the last three columns. We take these results based on the weighted mean and the polynomial fit as our fiducial results. For each line, the ``other'' sample excludes the 15 objects reported in \citet{Shen_etal_2016a}, and the ``high-$L$'' and ``low-$L$'' samples are the full sample divided at the median luminosity (for \halpha\ and \hbeta\ only). The sample-averaged redshift, (host-corrected) quasar continuum luminosity and line widths (denoted by ``$<>$'') are the weighted means calculated using the same weights as for the weighted mean lags. The measurement uncertainties of the average line widths are negligible and not reported here (see text). The last three columns show the results for the stacks performed in the rest frame of individual quasars.  
\end{tablenotes}
\end{table*}